\documentclass[review,nonatbib]{elsarticle}

\usepackage{lineno,hyperref}

\usepackage[nolist]{acronym} 			
\usepackage{booktabs}					
\usepackage{tabularx}					
\usepackage{amsmath, amssymb,amsthm} 	
\usepackage{subfig}						
\usepackage{tikz}						
\usetikzlibrary{trees}					
\usepackage{comment}					
\usepackage[official]{eurosym}			
\usepackage[at]{easylist}               

\makeatletter
\let\c@author\relax
\makeatother
\usepackage[
    backend=biber,
    style=numeric-comp,
    sorting=none,
    doi=true,
    url=true,
    isbn=false,
    minnames=1,
    maxnames=999,
    giveninits=true
    ]{biblatex}
\DeclareFieldFormat[article,misc,inbook,book]{citetitle}{#1}
\DeclareFieldFormat[article,misc,inbook,book]{title}{#1} 

\DeclareBibliographyDriver{article}{%
  \printnames{author}%
  \setunit{\addspace}
  \printtext[parens]{\printfield{year}}
  \newunit\newblock
  \printfield{title}%
  \newunit\newblock
  \printfield{journaltitle}%
  \setunit{\addspace}
  \printfield{volume}%
  \printfield{number}%
  \newunit\newblock
  \printfield{pages}%
  \newunit\newblock
  \printfield{doi}%
}

\DeclareBibliographyDriver{misc}{%
  \printnames{author}
  \setunit{\addspace}
  \printtext[parens]{\printfield{year}}
  \newunit\newblock
  \printfield{title}
  \newunit\newblock
  \usebibmacro{doi+eprint+url}
}

\DeclareBibliographyDriver{techreport}{%
  \printnames{author}
  \setunit{\addspace}
  \printtext[parens]{\printfield{year}}
  \newunit\newblock
  \printfield{title}
  \newunit\newblock
  \usebibmacro{doi+eprint+url}
}

\usepackage{color,soul}

\usepackage{comment}                    
\usepackage{amsmath}
\usepackage{accents}

\usepackage[long]{optidef}
\usepackage{lscape}
\usepackage{subfig}
\usepackage[official]{eurosym}
\usepackage{url}

\usepackage{pdfpages}
\usepackage{xcolor}
\usepackage{graphicx}
\definecolor{palevioletred}{HTML}{DB7093}
\usetikzlibrary{arrows.meta}
\usepackage{float}

\makeatletter
\newcounter{manualsubequation}
\renewcommand{\themanualsubequation}{\alph{manualsubequation}}
\newcommand{\startsubequation}{%
  \setcounter{manualsubequation}{0}%
  \refstepcounter{equation}\ltx@label{manualsubeq\theequation}%
  \xdef\labelfor@subeq{manualsubeq\theequation}%
}
\newcommand{\tagsubequation}{%
  \stepcounter{manualsubequation}%
  \tag{\ref{\labelfor@subeq}\themanualsubequation}%
}
\let\subequationlabel\ltx@label
\makeatother

\journal{arXiv}

\usepackage{anysize}
\marginsize{2.5cm}{2.4cm}{1.5cm}{2cm}

\begin{document}

\begin{frontmatter}

\title{Multi-threshold time series analysis enables characterization of variable renewable energy droughts in Europe}



\author[A1,A2]{Martin Kittel\corref{MK}}
\ead{mkittel@diw.de}

\author[A1]{Wolf-Peter Schill}
\ead{wschill@diw.de}

\cortext[MK]{Corresponding author}
\address[A1] {DIW Berlin, Department of Energy, Transportation, Environment, Mohrenstra{\ss}e 58, 10117 Berlin, Germany}
\address[A2] {Technical University Berlin, Digital Transformation in Energy Systems, Einsteinufer 25 (TA 8), 10587 Berlin, Germany}

\begin{abstract}

Variable renewable energy droughts, so called Dunkelflaute events, emerge as a challenge for climate-neutral energy systems based on variable renewables. Here we characterize European drought events for on- and offshore wind power, solar photovoltaics, and renewable technology portfolios, using 38 historic weather years and an advanced identification method. Their characteristics heavily depend on the chosen drought threshold, questioning the usefulness of single-threshold analyses. Applying a multi-threshold framework, we quantify how the complementarity of wind and solar power temporally and spatially alleviates drought frequency, return periods, duration, and severity within (portfolio effect) and across countries (balancing effect). We identify the most extreme droughts, which drive major discharging periods of long-duration storage in a fully renewable European energy system, based on a policy-relevant decarbonization scenario. Such events comprise sequences of shorter droughts of varying severity. The most extreme event occurred in winter 1996/97 and lasted 55~days in an idealized, perfectly interconnected setting. The average renewable availability during this period was still 47\% of its long-run mean. System planners must consider such events when planning for storage and other flexibility technologies. Methodologically, we conclude that using arbitrary single calendar years is not suitable for modeling weather-resilient energy scenarios.

\vspace{.5cm}


\begin{acronym}
 \acro{CSP}{Centralized Solar Power}
 \acro{ENTSO-e}{European Network of Transmission System Operators for Electricity}
 \acro{DIETER}{Dispatch and Investment Evaluation Tool with Endogenous Renewables}
 \acro{MBT}{Mean Below Threshold}
 \acro{NTC}{Net Transfer Capacity}
 \acro{PRL}{Positive residual load}
 \acro{PV}{photovoltaics}
 \acro{RoR}{Run-of-River}
 \acro{TYNDP}{Ten Year Network Development Plan}
 \acro{VRE}{variable renewable energy}
 \acro{VMBT}{variable-duration mean below threshold}
\end{acronym}

\end{abstract}


\end{frontmatter}

\section{Introduction}

Mitigating climate change requires deep reductions of greenhouse gas emissions that originate from fossil fuel use \cite{lee_ipcc_2023}. Nuclear energy or fossil fuels in combination with carbon capture and storage are options for this  \cite{sepulveda_role_2018}, but these firm low-carbon technologies face economic challenges and political controversies in many countries. Another major strategy for achieving climate neutrality is shifting energy supply to renewable energy sources, which are globally on the rise as they become increasingly cost-competitive \cite{irena_renewable_2025}. However, the potential for dispatchable renewable energy sources such as hydro power, geothermal, or bioenergy is limited in most countries. Hence, future renewable energy systems are likely to heavily rely on variable wind and solar power \cite{iea_net_2023,way_empirically_2022}. As these \ac{VRE} sources may become the most relevant primary energy sources in many countries \cite{jacobson_100_2017,brown_response_2018,child_flexible_2019,pickering_diversity_2022,wang_accelerating_2023}, energy systems are increasingly exposed to weather variability. In turn, spatial and temporal system flexibility is increasingly needed to match the supply of \ac{VRE} sources with demand. This includes geographical balancing via transmission, different types of energy storage, and demand response \cite{denholm_grid_2011,rasmussen_storage_2012,schlachtberger_benefits_2017,roth_geographical_2023,he_sector_2021,hunter_techno-economic_2021,tong_geophysical_2021}.

While future energy systems with high \ac{VRE} shares have to deal with variability on different timescales, extreme periods of low wind and solar availability emerge as a key challenge for realizing such renewable energy systems. Often referred to as ``Dunkelflaute'', these \ac{VRE} drought periods are characterized by long-lasting and substantial shortages of wind and solar energy \cite{raynaud_energy_2018,kittel_measuring_2024} and may cover large geographical areas. 
If the availability of firm low-carbon generation technologies is low, dealing with \ac{VRE} drought events necessitates the use of long-duration storage as well as other temporal and spatial flexibility options \cite{dowling_role_2020,sepulveda_design_2021}. Hence, understanding the spatial and temporal characteristics of \ac{VRE} drought events is crucial for weather-resilient energy system planning and energy policy as well as for the design of energy markets and support instruments for generation and flexibility technologies. This includes questions as to how frequent, how long, and how severe such periods are, and assessing their spatial and temporal correlation across large-scale interconnected energy systems.

Literature from different fields contributes to \ac{VRE} drought analysis. Wind droughts in the United Kingdom are well-studied \cite{cannon_using_2015,potisomporn_spatial_2022,potisomporn_evaluating_2023,abdelaziz_assessing_2024}, focusing on frequency-duration distributions, return periods as well as spatial and temporal correlations of historic or future on- and offshore wind droughts. Similar analyses for wind power have been conducted for Ireland \cite{leahy_persistence_2013}, the North Sea \cite{patlakas_low_2017}, Germany \cite{ohlendorf_frequency_2020}, or, analyzing deviations from climatological means, globally \cite{antonini_identification_2024}. Additionally, research interest in drought patterns of policy-relevant portfolios comprising wind and solar \ac{PV} is growing. Historic, future, and synthetic weather data of various world regions have been analyzed, such as Europe \cite{raynaud_energy_2018,kies_critical_2021,kapica_potential_2024,hu_implications_2023,breyer_reflecting_2022,van_der_most_temporally_2024,van_duinen_meteorological_2025}, the U.S. \cite{rinaldi_wind_2021,bracken_standardized_2024,bracken_seasonal_2025}, China \cite{li_renewable_2024,lei_frequency_2024}, India \cite{gangopadhyay_role_2022}, Germany \cite{kaspar_climatological_2019,mockert_meteorological_2023}, Hungary \cite{mayer_probabilistic_2023}, Japan \cite{ohba_climatology_2022}, and Australia \cite{richardson_climate_2023}. Besides regional and seasonal variations for single technologies and \ac{VRE} portfolios, a general finding is that combining wind and solar within regions, as well as considering balancing across regions, can mitigate drought characteristics.

Different methodological approaches have been used for renewable drought identification \cite{ohlendorf_frequency_2020,potisomporn_extreme_2024,kittel_measuring_2024,li_renewable_2024,biewald_evaluation_2025, wilczak_wind_2025}. Droughts can be defined as periods of consecutive time steps with renewable availability below a certain drought threshold, either with fixed \cite{raynaud_energy_2018,rinaldi_wind_2021,kies_critical_2021,gangopadhyay_role_2022,kapica_potential_2024,kaspar_climatological_2019,ohba_climatology_2022} or variable duration \cite{potisomporn_extreme_2024},
using various identification methods. For instance, these periods can be identified by searching for an availability constantly below the threshold \cite{potisomporn_spatial_2022,potisomporn_evaluating_2023,abdelaziz_assessing_2024,ohlendorf_frequency_2020,kies_critical_2021,breyer_reflecting_2022,kaspar_climatological_2019,mayer_probabilistic_2023,ohba_climatology_2022}, a mean availability over a certain averaging interval below the threshold \cite{abdelaziz_assessing_2024,leahy_persistence_2013,patlakas_low_2017,ohlendorf_frequency_2020,raynaud_energy_2018,rinaldi_wind_2021,gangopadhyay_role_2022,kapica_potential_2024,mockert_meteorological_2023}, or the deviation of the availability from a drought threshold \cite{potisomporn_extreme_2024} or its climatological mean \cite{antonini_identification_2024,stoop_climatological_2024}. Drought thresholds are either exogenously and presumably arbitrarily set \cite{cannon_using_2015,potisomporn_extreme_2024,abdelaziz_assessing_2024,leahy_persistence_2013,patlakas_low_2017,ohlendorf_frequency_2020,kies_critical_2021,kaspar_climatological_2019,mockert_meteorological_2023,mayer_probabilistic_2023,ohba_climatology_2022}, or derived from the data analyzed, such as a fraction of the time-invariant mean \cite{raynaud_energy_2018,rinaldi_wind_2021,kapica_potential_2024} or maximum \cite{breyer_reflecting_2022} availability, or of its time-variant climatological mean \cite{gangopadhyay_role_2022,antonini_identification_2024,stoop_climatological_2024}. Other researchers have proposed using copulas to construct joint cumulative distribution functions of drought severity and duration \cite{otero_copula-based_2022} or standardized indices to monitor and compare drought events \cite{allen_standardised_2023}. As each of these approaches has strengths and weaknesses, no standard drought identification method has yet emerged in the literature \cite{kittel_measuring_2024}.

Here we analyze and compare \ac{VRE} drought events for single renewable technologies and for a fully renewable European electricity system, using \ac{VRE} portfolio assumptions from a policy-relevant decarbonization scenario from European transmission grid operators. We do so for 33~individual European countries (EU27, the United Kingdom, Norway, Switzerland, and the Western Balkans), and for a pan-European ``copperplate'' scenario with perfect interconnection across all countries that allows for unconstrained geographical balancing. The latter represents an idealized scenario that allows identifying the upper bound of the value of geographical balancing for mitigating renewable energy droughts. We draw on a large dataset of historical \ac{VRE} availability factor time series covering 38~years \cite{de_felice_entso-e_2022}. We use an advanced, open-source algorithm that fully captures unique drought periods, properly accounts for brief periods of higher renewable availability within longer drought events, and avoids arbitrary threshold choices \cite{kittel_measuring_2024}. 
Note that the question what mix of storage and other flexibility technologies optimally facilitates weather-resilient renewable energy system is not in the focus of our analysis. This has to be investigated using computationally intensive numerical energy system models. However, the methods and findings presented here can help to identify relevant weather years that should be used in such model analyses.

Our results quantify how the spatial and temporal complementarity of wind and solar \ac{PV} alleviates \ac{VRE} droughts within European countries, giving rise to a technology \textit{portfolio effect}. Averaged over all thresholds and countries, the maximum drought duration of a renewable technology portfolio that combines solar \ac{PV} as well as on- and offshore wind power decreases by 64\%, 52\%, or 47\% compared to standalone \ac{PV}, onshore wind, or offshore wind droughts. We also demonstrate how European integration can further mitigate \ac{VRE} droughts by leveraging spatially complementary \ac{VRE} availability profiles across regions, signified by a \textit{balancing effect}. Considering unconstrained geographical balancing across all countries, the longest technology portfolio drought shortens by 65\%. We further show that drought characteristics strongly depend on the chosen drought threshold and that single-threshold analyses lead to an incomplete characterization of extreme drought patterns. Using a multi-threshold analysis, we illustrate that the most extreme \ac{VRE} events manifest as sequential droughts with varying severity, affecting multiple countries simultaneously to different extents. We introduce a drought mass indicator to identify the most extreme droughts and illustrate how these determine long-duration storage needs in a fully renewable European energy system. Considering Germany alone, the most extreme event occurred in the winter of 1995/96 and lasted 109~days. Under the idealized assumption of unconstrained geographical balancing across Europe, the most extreme storage-defining event was substantially shorter and lasted 55~days, occurring in the winter of 1996/97. Such extreme droughts may occur at the turn of years, suggesting that planning horizons based on single calendar years are inappropriate for modeling weather-resilient future energy systems.

\section{Results}

We define a renewable energy drought as a period during which the moving average of the hourly \ac{VRE} availability factor, also referred to as capacity factor, is below a given qualification threshold~\cite{kittel_measuring_2024}. The search algorithm used here iteratively decreases the moving average window, which initially identifies the longest-lasting droughts, followed by progressively shorter ones. As there is no consensus in the literature on appropriate drought thresholds, we apply a large number of thresholds to each investigated region and technology, ranging in 5\% increments from 10\% to 100\% of the mean renewable availability across all investigated years \cite{kittel_measuring_2024}. While low thresholds identify severe droughts, higher thresholds increasingly include more moderate drought events. By scaling drought thresholds relative to the mean availability factor of all hours in the data of each region-technology setting (henceforth referred to as ``relative thresholds''), we account for regional and technology-specific differences in generation potentials. This enables meaningful cross-regional and cross-technological comparisons. Section~\ref{ssec:methods_vreda} details the renewable energy drought definition and identification method. To give an example, an identified wind power drought with a 24-hour duration at a threshold $\tau_{0.1}$ should be interpreted such that less than 10\% of the long-run average electricity generation of wind power is available on average during a 24-hour period.

In the main part of the paper, we illustrate selected results for a limited set of countries. The \ref{sec:sup_inf} provides additional results and information on the thresholds used. Interactive, high-resolution versions of all graphs that allow for retrieving concrete values for each of the illustrated 3D-plots as well as an additional animated graph are available online \cite{kittel_high_2025}.

\subsection{Complex renewable energy drought dynamics and complementarities across regions and technologies}\label{ssec:drought_patterns}

\begin{figure}[ht]
\centering
\noindent\includegraphics[width=\linewidth,keepaspectratio]{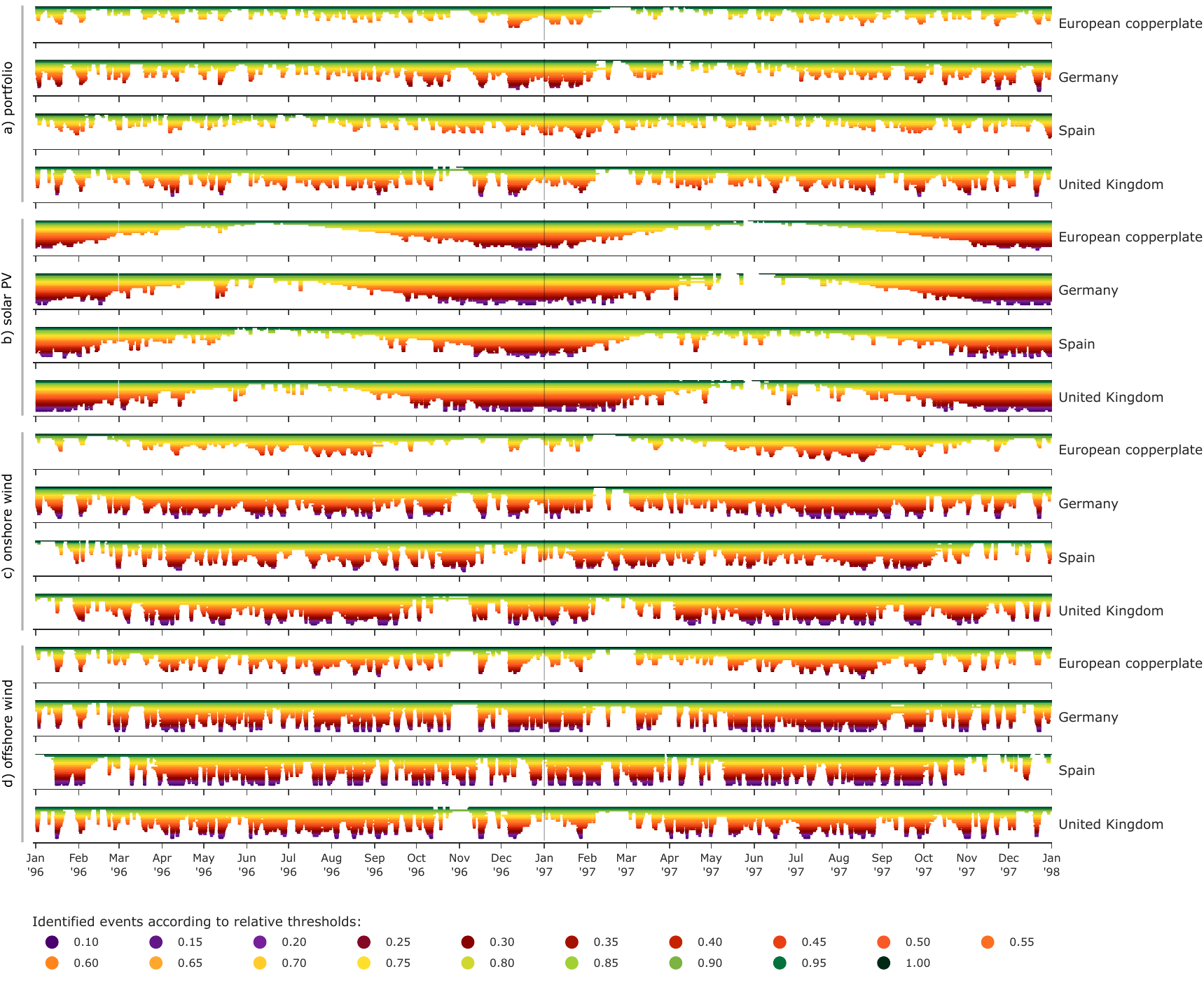}
\caption{Drought patterns in 1996 and 1997 for selected regions, corresponding to the following technology settings: a) renewable technology portfolio combining solar \ac{PV} as well as on- and offshore wind, b) solar \ac{PV}, c) onshore wind, and d) offshore wind. \\ Identified drought patterns in 1996 and 1997 for all investigated relative thresholds $\tau_i$ with $i \in [0.1,..., 1]$ and selected regions. The dates shown on the bottom axis correspond to all technology-region-specific panels. For each panel, the stacked horizontal bands indicate drought occurrences for the color-coded threshold of a particular region. To illustrate persistent patterns, only droughts lasting longer than one day are displayed. Figure~\ref{fig:figure_SI2} focuses on longer-lasting events by illustrating respective patterns for droughts lasting at least one week.}
\label{fig:figure_1}
\end{figure}

Figure~\ref{fig:figure_1} shows renewable energy droughts that last longer than one day for the years 1996 and 1997, which emerge to be particularly critical (see Section~\ref{ssec:drought_mass}). Figure~\ref{fig:figure_1} combines these events for all employed relative thresholds, for selected countries from southern, central and northern Europe, as well as for the pan-European copperplate scenario. These countries represent some of the largest electricity markets in Europe, each with distinct yet complementary wind and solar generation profiles. The pan-European copperplate scenario combines respective profiles from all 33 countries in our data set, assuming unconstrained geographical balancing. Such an idealized scenario is unlikely to be ever achieved in practice, as it would require expanding European grids on a massive scale. However, the copperplate scenario allows identifying the upper bound of the value of European integration to mitigate renewable droughts.
The multi-threshold illustration shows the variety of renewable drought patterns in Europe across regions and technologies even within the same time period. It further indicates that the identified drought duration is highly sensitive to the underlying color-coded threshold. In general, single-threshold analyses using low thresholds can detect extreme short-duration droughts, which may occur in isolation or adjacent to high-availability periods (white spaces in the Figure). Higher thresholds increasingly identify longer-lasting events. These potentially include consecutive shorter periods of extreme renewable shortage or surplus, which are smoothed by the averaging mechanics of the identification algorithm (see Section~\ref{ssec:methods_vreda}). The multi-threshold perspective demonstrates that longer-lasting events identified by higher thresholds may encompass multiple severe droughts detected at lower thresholds.

Solar \ac{PV} events identified by moderate or high relative thresholds can last several months, indicating low solar availability during winter in Europe (Figure~\ref{fig:figure_1}, panel b). Lower thresholds identify very severe \ac{PV} droughts driven by low seasonal availability in combination with factors such as persistent cloud or snow coverage, such as in winter 1996. In summer, only rare and isolated \ac{PV} droughts of lower severity occur that are longer than one day. These seasonal \ac{PV} characteristics can generally be found across all years and are often more pronounced in Northern Europe, e.g.,~observable in winter 2011/12 (compare Germany or United Kingdom to Spain in Figure~\ref{fig:figure_SI3}).

In contrast, severe long-lasting onshore and offshore wind droughts can occur throughout the year but tend to be more frequent in summer (Figure~\ref{fig:figure_1}, panels c and d), which is in line with general wind seasonality \cite{kaspar_climatological_2019}. The multi-threshold perspective further reveals that severe wind droughts identified by lower thresholds can be sequential and may occur within contiguous below-average wind periods detected at higher thresholds. The latter may last up to several months. While such long-lasting events usually occur in summer (compare Figures~\ref{fig:figure_SI3}-\ref{fig:figure_SI5}), they may also take place in winter in some countries (e.g., in 1996/97, Figure~\ref{fig:figure_1}). On- and offshore droughts are generally correlated, but may differ in severity (Figure~\ref{fig:figure_1}, panels c and d).

The multi-threshold illustration also reveals a technology \textit{portfolio effect} due to complementary wind and solar power droughts. Typically, high solar availability offsets wind droughts in summer, while wind power often compensates for the low solar \ac{PV} availability in winter. This results not only in briefer but also less severe renewable portfolio droughts, i.e.,~higher thresholds apply in the same periods compared to single technologies (compare panel a of Figure~\ref{fig:figure_1} with panels b-d). However, if severe and persistent wind and \ac{PV} droughts coincide, pronounced long-lasting compound droughts can occur even for renewable portfolios, as seen for instance in winter 1996/97.

Figure~\ref{fig:figure_1} also illustrates a geographical \textit{balancing effect}. Assuming perfect interconnection across Europe, geographical balancing can substantially mitigate the severity and duration of renewable energy droughts. This can be seen when comparing the European copperplate scenario (top row in each panel) to the corresponding isolated-countries cases in Figure~\ref{fig:figure_1}. This \textit{balancing effect} is particularly pronounced for onshore wind, as onshore wind droughts tend to occur at different times across space, which allows for spatial smoothing across all 33~considered countries. The effect is less pronounced for offshore wind, which can only be deployed in countries with sea access. This limits the spatial extent over which balancing is possible. It is even less pronounced for solar \ac{PV} because \ac{PV} droughts are heavily driven by solar seasonality, which affects most countries simultaneously.

Geographical balancing also reduces the severity and duration of technology portfolio drought events as solar seasonality is lower in Southern Europe and wind droughts do not occur simultaneously across countries. Severe and long-lasting portfolio droughts may arise in single countries and also across Europe at the beginning of a calendar year, (e.g.,~observable in 2012, Figure~\ref{fig:figure_SI3}), at the end of a year (e.g.,~1982, Figure~\ref{fig:figure_SI4}), or across the turn of years (e.g.,~in winter 1996/97, Figure~\ref{fig:figure_1}). This can have important consequences for weather-resilient energy modeling (see Section~\ref{sec:discussion}). In some years, winter portfolio droughts are much less pronounced (e.g.,~2013/14, Figure~\ref{fig:figure_SI5}). Countries relying heavily on wind power, such as the United Kingdom, may also experience very severe events in summer.

While this section highlights important technology-specific and regional characteristics of renewable energy droughts in Europe drawing on specific weather years, it necessarily remains selective. Figure~\ref{fig:figure_SI6} shows the threshold-specific distribution of all identified droughts across all years in the data. 
These distributions are highly sensitive to the chosen threshold. While the majority of droughts are relatively short, particularly for lower drought thresholds, higher thresholds find increasingly long-lasting events across all weather years. Further, the distributions confirm the \textit{portfolio effect} from combining complementary wind and solar \ac{PV} profiles into a technology portfolio discussed above, both in terms of mean and maximum duration.
In the following, we systematically analyze both shorter and very long-lasting renewable droughts for all technologies, regions, and weather years in our data. In doing so, we also connect to and extend analyses based on single thresholds which have been prevalent in the literature so far.

\subsection{Frequency-duration distribution: short drought events occur much more often than longer ones}\label{ssec:frequencies}

\begin{figure}[htbp]
\centering
\noindent\includegraphics[width=\linewidth,height=\textheight, keepaspectratio]{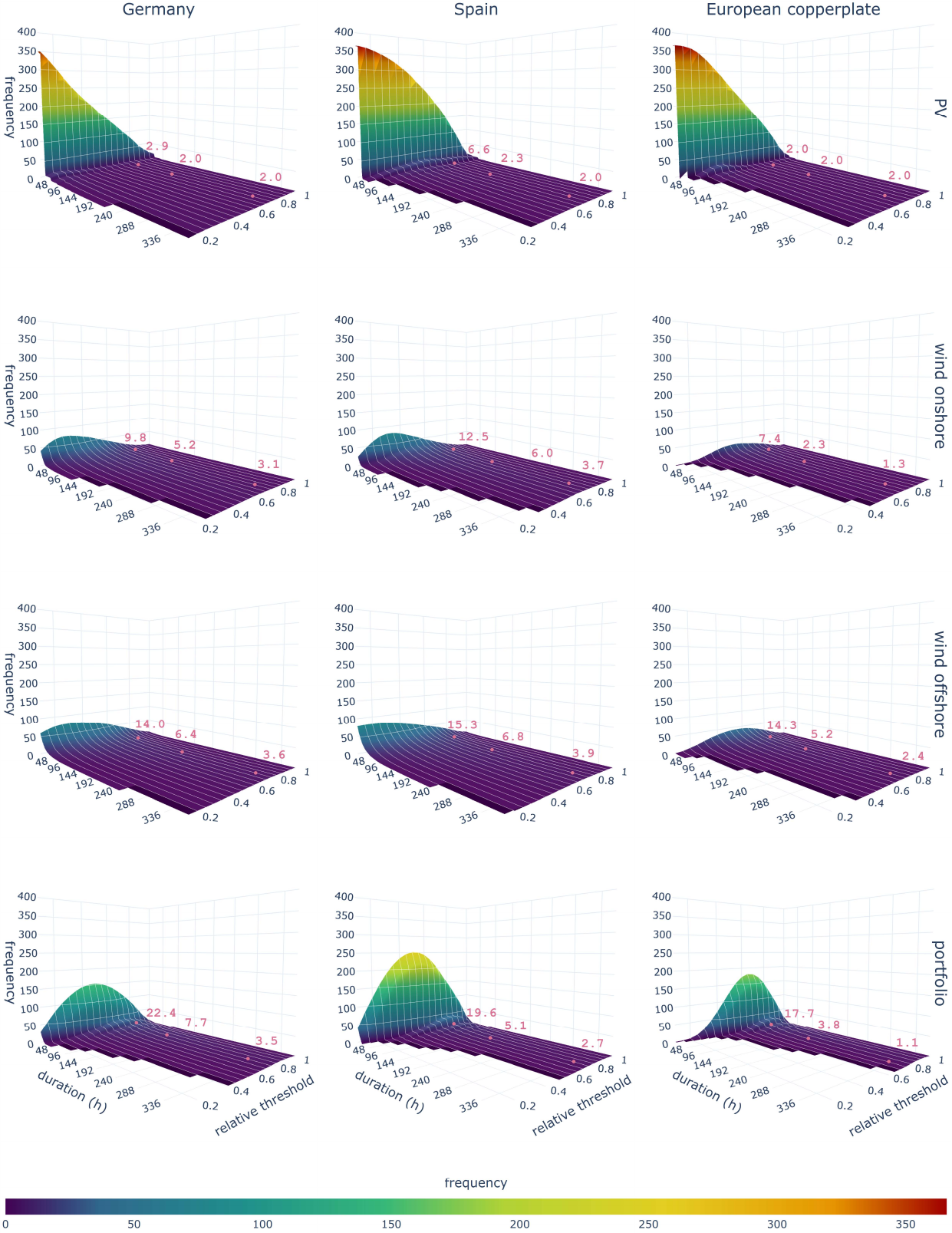}
\caption{Example of frequency-duration distributions of drought events. \\ Cumulative frequency-duration distributions of drought events across all investigated relative drought thresholds $\tau_i$ with $i \in [0.1,..., 1]$, sorting the yearly frequencies of all events that are at least as long as a given duration. White space indicates the absence of droughts for given thresholds in the data. The contour lines represent the threshold-specific yearly frequency. For illustration, the distributions are truncated at 360 hours, i.e., they show events with a maximum duration of just above two weeks. Yearly frequencies of events lasting at least two days, one week, and a fortnight are marked for a relative threshold of $\tau_{0.75}$. An interactive version of this Figure which allows zooming and rotating is available online \cite{kittel_high_2025}.}
\label{fig:figure_2}
\end{figure}

A cumulative frequency-duration distribution shows how often \ac{VRE} droughts that lasted at least a certain duration occurred on average per year in the investigated data. Figure~\ref{fig:figure_2} illustrates these distributions across all investigated relative thresholds for selected regions. For a given threshold, the frequencies of all events that are at least as long as a given duration are shown in ascending order on the vertical axis. For instance, at a relative threshold $\tau_{0.75}$, Germany experienced on average 9.8 onshore wind droughts lasting at least 48~hours (two days), 5.2 lasting at least 168~hours (one week), and 3.1 lasting at least 336~hours (two weeks) per year. For comparability, Figure~\ref{fig:figure_2} uses the same frequency-axis range across all technologies and regions. Figure~\ref{fig:figure_SI7} focuses on on- and offshore wind droughts using an adjusted frequency-axis range. Also note that the duration axis in Figure~\ref{fig:figure_2} is truncated at 360 hours, so it focuses on the occurrence of shorter-duration events. These are particularly relevant for the operation of shorter-duration flexibility options in the power sector, e.g.,~battery or hydro reservoir storage. Figure~\ref{fig:figure_SI8} shows longer-lasting drought exceeding 360 hours to up to one full year, which may drive the need for long-duration electricity storage or other firm low-carbon technologies. Additionally, the following sections analyze such extreme periods in depth.

In general, very short drought events occur much more often than longer ones for every given threshold. Events with very short duration reflect typical diurnal renewable variability, and should not be interpreted as exceptional periods of low availability. Long droughts lasting more than a week are infrequent. Importantly, the threshold choice substantially impacts the frequency of drought events. For lower thresholds, we generally find fewer events compared to higher thresholds. An exception is the left-hand side of the frequency-duration distributions shown in Figure~\ref{fig:figure_2}. For very short drought durations, we find the largest number of events for relatively low thresholds, and a decreasing frequency of droughts for higher thresholds. This is because higher thresholds tend to identify fewer, yet longer-lasting droughts, which combine multiple shorter events that are counted as individual events for lower thresholds.

The cumulative frequency-duration distribution substantially differs between wind and solar power. For \ac{PV} and low thresholds, we identify many events that last less than one day. They reflect the typical diurnal fluctuations of solar \ac{PV}, with zero production at night-time \cite{ueckerdt_analyzing_2015}. Increasing thresholds lead to fewer \ac{PV} drought events, as many night-time periods of zero electricity generation merge into multi-day events. When the threshold approaches $\tau_{1.0}$, the identified droughts increasingly reflect a strong solar seasonality in Europe, with lower availability in winter than in summer. For on- and offshore wind power, droughts lasting less than one day occur less often. This is because wind power does not have regular non-availability at night-time as solar \ac{PV}, but fluctuates less regularly. In turn, events that last longer than one day are more frequent, reflecting a typical multi-day variability of wind power \cite{apt_spectrum_2007,ueckerdt_analyzing_2015}.

The frequency-duration distribution of \ac{VRE} portfolio droughts also illustrates the \textit{portfolio effect}, i.e.,~droughts identified by lower thresholds (see white spaces in the portfolio row in Figure~\ref{fig:figure_2} are less frequent when wind and solar \ac{PV} are combined due to complementary wind and solar availability. Hardly any \ac{VRE} portfolio droughts arise for lower thresholds. For higher thresholds, their frequency still is much lower than for single technologies (compare highlighted frequencies in each column in Figure~\ref{fig:figure_2}). 

Figure~\ref{fig:figure_2} also illustrates the geographical \textit{balancing effect}, as droughts occur less often in an ideal European interconnection than in single countries. This is true for both single \ac{VRE} technologies and portfolios. The reason for this is that periods of low solar and wind availability are not perfectly correlated across Europe \cite{roth_geographical_2023}. For example, assuming a relative threshold $\tau_{0.75}$, there have been on average approximately eight and five \ac{VRE} portfolio droughts per year in Germany and Spain that lasted at least one week, respectively, while less than four of such events occurred in Europe. For on- and offshore wind, the \textit{balancing effect} mitigates drought frequency, particularly for drought events identified by lower thresholds, i.e.,~more severe wind droughts. This effect is visualized by the additional white space in Figure~\ref{fig:figure_2} and the frequency-duration distributions in Figure~\ref{fig:figure_SI7} when in the European copperplate scenario compared to Germany or Spain, respectively. 

In the \ref{sec:sup_inf}, we provide additional results on the seasonality of drought patterns for Europe under the assumption of perfect interconnection. For low thresholds and events lasting longer than a few days, \ac{PV} droughts are more frequent in winter, while wind droughts are more frequent in summer (Figure~\ref{fig:figure_SI9}).

\subsection{Return periods: most extreme events occur rarely}\label{ssec:sup_inf_return_periods}

\begin{figure}[htbp]
\centering
\noindent\includegraphics[width=\linewidth, keepaspectratio]{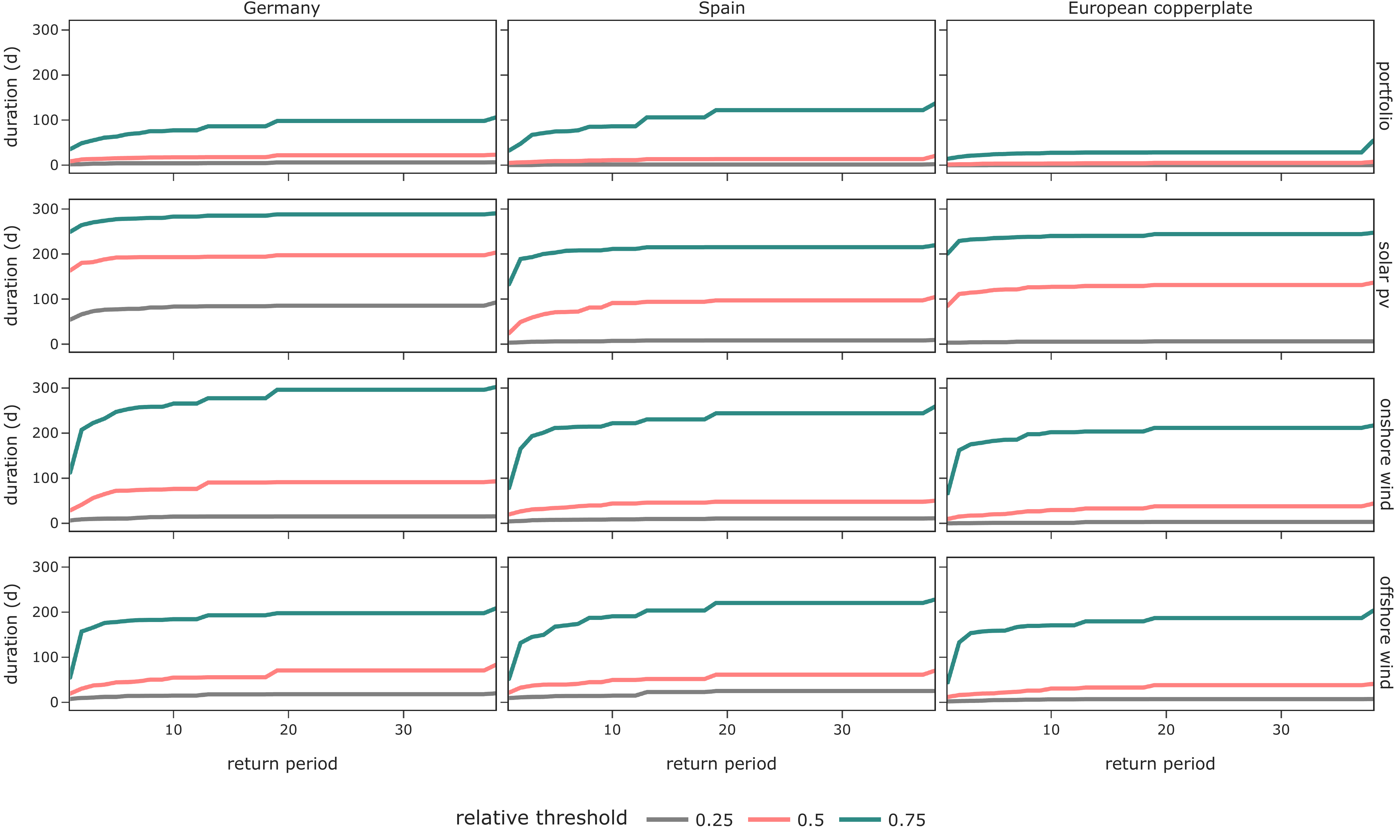}
\caption{Example of return period curves of extreme drought events. \\ Return period curves of extreme drought events with an average annual frequency below 1 across selected relative thresholds $\tau_i$. Figure~\ref{fig:figure_SI10} provides a more comprehensive perspective on return periods across all thresholds.}
\label{fig:figure_3}
\end{figure}

Return periods of \ac{VRE} droughts are given by the reciprocal of average yearly frequencies. 
Return-period curves for extreme events indicate the longest event duration that can, on average, be expected to reoccur after a given number of years (Figure~\ref{fig:figure_3}). Note that the return period monotonically increases and the maximum return period is limited by the 38 years of data that we investigate. In \ref{ssec:sup_inf_return_period}, we provide more detailed, three-dimensional return-period visualizations.

For each threshold, the maximum event duration increases with higher return periods, 
but is highly sensitive to the underlying threshold. For any given return period, higher thresholds lead to events with longer maximum durations, plateauing at 365 days for very high thresholds (Figure~\ref{fig:figure_SI10}). Note that such high durations indicate below-average wind and solar years and should not be mistaken for actual drought events. The return period-duration curves vary substantially between technologies and countries. For example, the maximum offshore wind drought for a relative threshold of $\tau_{0.75}$ in Germany that can be expected to return every 20 years lasts 198~days and is much shorter than the corresponding \ac{PV} (289~days) and onshore wind droughts (297~days). By comparison, in Spain, such 20-year return period droughts are shorter for \ac{PV} (216~days) and onshore wind (245~days) but longer for offshore wind (221~days).

When combined in a technology portfolio, the maximum duration decreases substantially for each return period in both Germany and Spain and also in the European copperplate, which is another instance of the \textit{portfolio effect}. The \textit{balancing effect} is also visible: with perfect interconnection, the maximum drought duration further decreases due to a limited temporal correlation of wind droughts across Europe, especially for the renewable technology portfolio. In the case of solar \ac{PV}, higher solar availability in South Europe during winter balances lower availability in Northern Europe.

\subsection{Maximum drought duration strongly depends on threshold and differs across years and seasons}\label{ssec:max_duration}

\begin{figure}[htbp]
\centering
\noindent\includegraphics[width=\linewidth,height=\textheight, keepaspectratio]{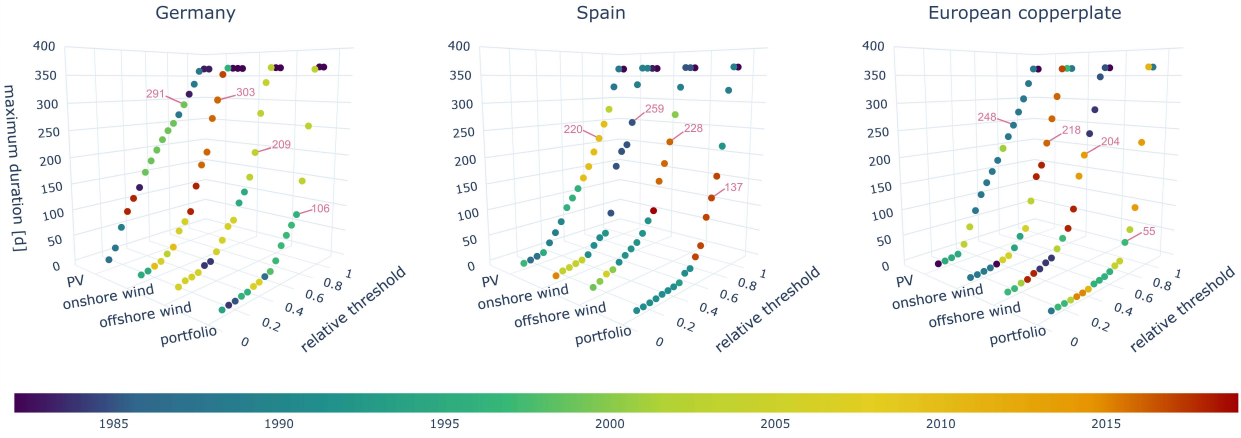}
\caption{Maximum duration of single drought events. \\ Most extreme duration of single drought events across all investigated years for given relative thresholds $\tau_i$ with $i \in [0.1,..., 1]$. The year with the most extreme event duration varies across thresholds. The marked numbers indicate the longest drought durations found for a relative threshold of $\tau_{0.75}$.}
\label{fig:figure_4}
\end{figure}

Figure~\ref{fig:figure_4} shows the longest droughts obtained from the data across all investigated years for different relative thresholds and technologies. A general finding is that the maximum drought duration strongly depends on the threshold. For very low thresholds around $\tau_{0.1}$, the maximum duration of all droughts in the data is very short. For higher thresholds approaching $\tau_{1.0}$, the maximum drought duration grows strongly and eventually reaches a plateau of 365~days, i.e.,~a full year. This is because at very high thresholds, the search algorithm identifies years with below-average renewable availability as drought events. Using very high thresholds close to $\tau_{1.0}$ thus appears not to be meaningful for identifying droughts relevant for energy system operations within a given year.

For medium thresholds, we find that both a technology portfolio and geographical balancing strongly decrease the maximum drought duration. For example, for a relative threshold of $\tau_{0.75}$, the longest \ac{PV} drought in Germany lasts 291~days, while the maximum on- and offshore wind droughts are 303 and 209 days, respectively. In the case of a combined \ac{VRE} portfolio, the respective maximum drought in Germany decreases to 106 days due to the technology \textit{portfolio effect}. If, on top, perfect interconnection in Europe is considered, the duration of the longest portfolio event for the same thresh  old further declines to 55~days due to the geographical \textit{balancing effect}. As visible in Figure~\ref{fig:figure_4}, the balancing effect is particularly pronounced for medium thresholds. This is because maximum drought durations are very short for very small thresholds close to 0 even without interconnection, and very long for large thresholds approaching 1 even with perfect interconnection. Accordingly, the balancing effect cannot play out for such extreme thresholds.

Especially for medium thresholds, the maximum yearly drought duration strongly depends on the weather year (Figure~\ref{fig:figure_SI11}). For example, the longest \ac{VRE} portfolio drought in Germany obtained for a relative threshold of $\tau_{0.75}$ occurred in 1996 (106~days), while the shortest yearly maximum drought duration was in 2018 (22~days). Considering perfect interconnection across all countries and the same threshold, the longest and shortest portfolio droughts occurred in 1997 and 1999, respectively (55 and 8~days).

The most extreme droughts vary substantially across countries, both with respect to their duration and the year of occurrence (Figures~\ref{fig:figure_SI12} and \ref{fig:figure_SI13}). Importantly, the ranking of years varies with the drought threshold (Figure~\ref{fig:figure_4}). For instance, the longest \ac{VRE} portfolio drought in Germany for a threshold of $\tau_{0.4}$ occurred in 2007 but for the thresholds $\tau_{0.6}$ and $\tau_{0.8}$ the most extreme droughts were in 1997 and 2003. The corresponding longest drought events in Spain (or in Europe, assuming perfect interconnection) occurred in 1991 (2005), 2017 (1996), and 2017 (2002). The spreads between the longest drought events in each year further increase with higher thresholds, before they decrease again approaching the threshold $\tau_{1.0}$ (Figure~\ref{fig:figure_SI14}). From these findings follows that any claim on a specific year being particularly relevant for energy modeling analyses, because it features a large renewable drought, should always be qualified with the respective threshold.

\begin{figure}[hp!]
\centering
\noindent\includegraphics[width=.97\linewidth,height=\textheight, keepaspectratio]{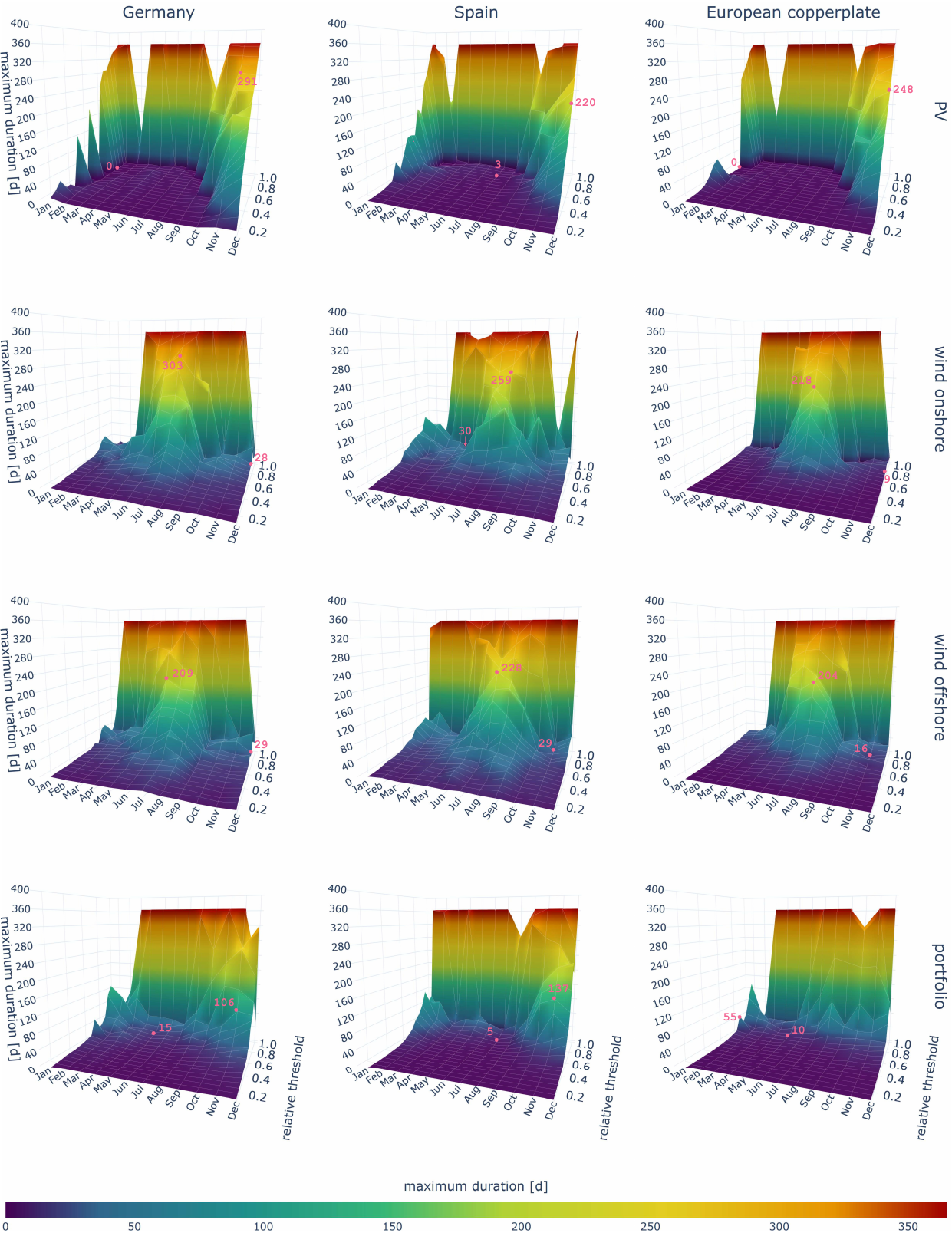}
\caption{Example of maximum duration of single drought events across months. \\ Most extreme duration of single drought events across all months in the data and all investigated thresholds $\tau_i$ with $i \in [0.1,..., 1]$. The contour lines represent the threshold-month-specific maximum duration. Note that the monthly assignment illustrates the median hour of respective droughts, while the duration of each event is plotted on the vertical axis. Events lasting longer than one month start (end) in previous (subsequent) months. The events with the highest and lowest duration across all months are marked for a threshold $\tau_{0.75}$.}
\label{fig:figure_5}
\end{figure}

The maximum drought duration further varies substantially across seasons and technologies. Solar \ac{PV} droughts primarily occur in fall and winter, with their median hour in November, December, or January (Figure~\ref{fig:figure_5}). The median hour represents the hour after half of the drought duration has passed. The maximum European \ac{PV} drought under unconstrained geographical balancing lasts 248~days with a median hour in December for a threshold of $\tau_{0.75}$. The longest \ac{PV} droughts with median hours in spring or summer months barely last longer than a few days, except for those identified by very high thresholds close or equal to $\tau_{1.0}$. In contrast, the longest European on- and offshore wind droughts mainly occur in spring and summer, with medians in July and maximum duration of 218 or 204~days, respectively. Maximum wind drought duration in winter months are much lower. In single countries, summer and winter wind droughts can be much longer (\textit{balancing effect}). The seasonal complementarity of wind and solar further mitigates combined technology portfolio droughts to a substantial extent (\textit{portfolio effect}). The longest-lasting droughts occur during winter months, which is a result of low seasonal \ac{PV} availability in combination with rare but severe concurrent wind droughts. The median hours of the longest renewable technology portfolio droughts in Germany (106~days) or Spain (137~days) occur in November, or, under the assumption of unconstrained balancing, in January (55~days). 

\subsection{Portfolio and balancing effects mitigate maximum drought durations}

\begin{figure}[htbp]
\centering
\noindent\includegraphics[width=.97\linewidth, keepaspectratio]{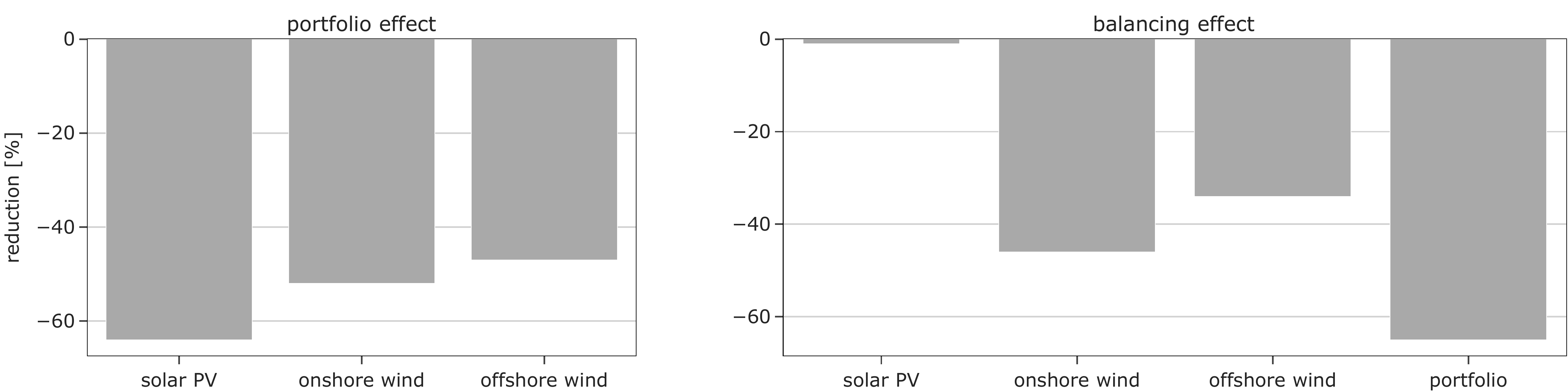}
\caption{Average portfolio and balancing effects. \\ Reduction of maximum drought durations for single renewable technologies compared to a renewable technology portfolio within a country (portfolio effect) and when considering unconstrained geographical balancing across all countries (balancing effect) averaged over all thresholds and countries.}
\label{fig:figure_6}
\end{figure}

Comparing weighted averages over all thresholds and countries, the maximum drought duration of a \ac{VRE} portfolio decreases by 64\%, 52\%, or 47\% compared to solar \ac{PV}, onshore wind, or offshore wind alone (Figure~\ref{fig:figure_6}). This \textit{portfolio effect} is more pronounced for \ac{PV} than for wind power, as onshore and offshore wind power complement both diurnal and seasonal shortages of \ac{PV}. In contrast, wind power does not incur diurnal shortages. Assuming a European copperplate (CP), the \textit{portfolio effect} (-80\%) for solar \ac{PV} is even stronger than for any isolated country (Table~\ref{tab:table_1}). This is because in a larger geographical area, there is more complementary wind power available to compensate for solar \ac{PV} shortages \cite{roth_geographical_2023}. In turn, the \textit{portfolio effect} is less pronounced for onshore wind power in the European interconnection (-45\%) as \ac{PV} availability is more homogeneously distributed. For offshore wind power, the \textit{portfolio effect} is higher (-70\%), as only a subset of countries features this technology, such that solar and wind onshore resources of all countries can compensate for extreme offshore droughts in a subset of countries.

Considering unconstrained geographical balancing across all countries, the longest \ac{PV}, onshore wind, or offshore wind droughts decrease by 1\%, 46\%, or 34\%, using weighted averages over all countries (Figure~\ref{fig:figure_6}). The very small average \textit{balancing effect} for \ac{PV}, however, conceals that there is a substantial north-south divide between countries with higher and lower solar irradiation (Table~\ref{tab:table_1}). For South European countries with abundant solar availability such as Spain, Portugal, Italy, or Greece, geographical balancing even increases solar \ac{PV} drought duration. In turn, geographical balancing alleviates maximum \ac{PV} drought durations in North European countries such as Germany or Scandinavia. In contrast, the \textit{balancing effect} on wind drought duration is more similarly distributed across countries, as maximum wind droughts rarely occur simultaneously throughout Europe. The \textit{balancing effect} is even more pronounced for \ac{VRE} technology portfolios, decreasing by 65\% on average. That is, geographical balancing not only helps to mitigate droughts for single renewable technologies, but even more so for a technology portfolio, due to complementary technology portfolios with different capacity mixes across countries.

\subsection{Drought mass: a multi-threshold metric to identify events relevant for long-duration electricity storage}\label{ssec:drought_mass}

Previous studies that use threshold-based identification methods typically use only a limited number of thresholds (e.g.,~\cite{leahy_persistence_2013,cannon_using_2015,patlakas_low_2017,kaspar_climatological_2019,ohlendorf_frequency_2020,kies_critical_2021,potisomporn_extreme_2024,potisomporn_extreme_2024,abdelaziz_assessing_2024,mockert_meteorological_2023,mayer_probabilistic_2023,ohba_climatology_2022}). However, the findings presented above indicate that drought characteristics strongly depend on the chosen threshold. To account for this, we introduce a multi-threshold ``drought mass'' metric to identify the most extreme portfolio drought events and show that it identifies events that are relevant for long-duration electricity storage in renewable European energy systems. In contrast to previous single-threshold approaches, the drought mass quantifies extreme events by integrating both drought duration and severity across a wide range of thresholds, thus properly accounting for the complex and diverse renewable drought patterns discussed in Section~\ref{ssec:drought_patterns}. It accumulates the number of drought hours identified at thresholds $\tau_{i \leq 0.75}$ that occur within contiguous events identified at $\tau_{0.75}$, i.e.,~excluding droughts found by very high thresholds ($\tau_{i > 0.75}$). The event with the highest score within a each pair of consecutive years in the data (i.e.,~1982-1983, 1983-1984, 1984-1985, etc.) is identified as the most extreme drought event within this time frame. Section~\ref{ssec:method_drought_mass} further elaborates on the drought mass mechanics.

\begin{figure}[htbp]
\centering
\noindent\includegraphics[width=\linewidth,keepaspectratio]{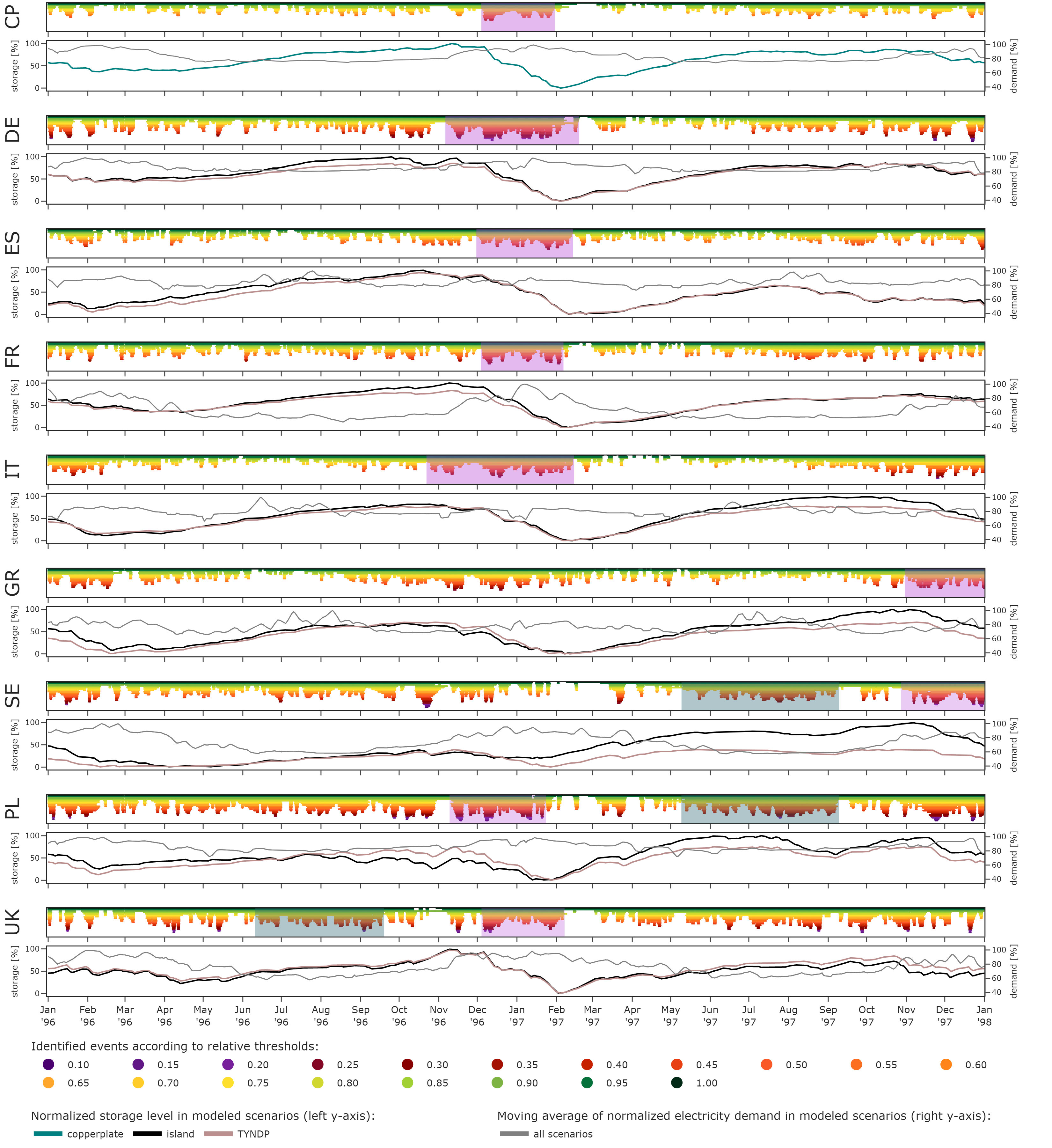}
\caption{Most extreme drought events in 1996/97 of selected countries.\\ Most extreme events identified by the drought mass metric occurring in winter (purple boxes). For countries in which the most extreme drought events occur in summer, these are additionally shown (teal boxes). The dates shown on the bottom axis correspond to all panels. For each region, portfolio drought occurrences lasting longer than one day for stacked, color-coded thresholds (upper panel) as well as normalized exogenous demand profiles and optimized storage levels across three modeled interconnection scenarios (lower panel) are displayed.}
\label{fig:figure_7}
\end{figure}

Figure~\ref{fig:figure_7} shows extreme portfolio drought patterns identified by the drought mass metric for the years 1996/97 and selected countries. The most extreme events, i.e.,~those with the highest drought mass, comprise sequences of shorter, but more severe droughts within contiguous well-below-average periods that may last up to several months. Events with the highest drought mass scores may occur in winter (purple boxes), potentially spanning across the turn of a calendar year, as in the case of the European copperplate in 1996/97. Severe drought mass events may also occur in summer (teal boxes) in countries that heavily rely on wind power such as the United Kingdom or Poland.

Note that renewable availability is not zero during these events, but it is low on average for a very prolonged period of time. For instance, the average availability factor of a renewable technology portfolio during the most extreme drought is 0.11 for the European copperplate scenario (55~days in winter of 1996/97), 0.07 in Germany (109~days in winter 1995/96), and 0.08 in Spain (131~days in winter 1988/89). For comparison, the average portfolio renewable availability factor over all hours in the data is 0.23 for the European copperplate, 0.19 for Germany, and 0.21 for Spain, respectively. That is, average renewable availability still amounts to 47\% of its long-run average during the most extreme drought event that defines long-duration storage needs in a fully renewable, perfectly interconnected European power sector. For Germany or Spain, the average availability during the most extreme drought amounts to 37\% or 35\%, respectively. Yet, within the largest drought mass events, we find more severe but shorter drought events with very low \ac{VRE} availability. For instance, during the extreme drought in the winter of 1996/97, events that lasted 17 and 18 days occurred in Germany with average availabilities of just over 0.05 identified by a threshold $\tau_{0.5}$, relating to 27\% and 29\% of the long-run average, respectively.

Peak electricity demand often depends on ambient temperature and occurs in winter in most European countries, except for the Mediterranean area. Accordingly, winter drought events that coincide with high electricity demand are particularly relevant. Using a stylized power sector model, we find that these compound events are a major driver for the use of long-duration electricity storage in all three interconnection states: countries as energy islands, a pan-European copperplate scenario, and policy-relevant interconnection levels between European countries according to the Ten-Year-Network-Development-Plan (``TYNDP scenario''). Sections~\ref{ssec:DIETER} and \ref{ssec:scenarios} elaborate further on the power sector model and interconnection scenarios. Figure~\ref{fig:figure_7} shows the optimal operation of long-duration electricity storage for different degrees of interconnection between countries as well as electricity demand patterns for the years 1996/97. The most severe identified droughts largely coincide with the periods of long-duration storage discharge, i.e.,~decreasing storage levels. This holds true for both the European copperplate scenario and several isolated countries in the island scenario. When considering policy-relevant interconnection in the TYNDP scenario, the storage level patterns do not structurally change but only show a level shift compared to the island scenarios. That is, drought patterns identified for isolated countries are a reasonable approximation of drought patterns in settings with more policy-relevant, limited interconnection with regard to long-duration storage operation.

In some countries, the most severe drought mass event does not coincide with the major discharging period of long-duration storage. For example, in Sweden, Poland, or the United Kingdom, the metric finds that the largest drought events occur in summer (gray boxes in the lowest three panels of Figure~\ref{fig:figure_7}). Long-duration storage usage, however, still coincides with the most extreme winter droughts, which are comparatively less severe. This is because electricity demand is lower during the more severe summer drought events identified in these countries. Consequently, summer droughts do not affect the system's ability to meet demand as much. In complementary model runs where we assume a flat electricity demand profile, i.e.,~abstracting from seasonal and diurnal variability, the most severe event measured by the drought mass metric perfectly coincides with storage-defining periods in all countries, including Sweden, Poland, and the United Kingdom (Figure~\ref{fig:figure_SI15}).

\begin{figure}[htbp]
\centering
\noindent\includegraphics[width=\linewidth,keepaspectratio]{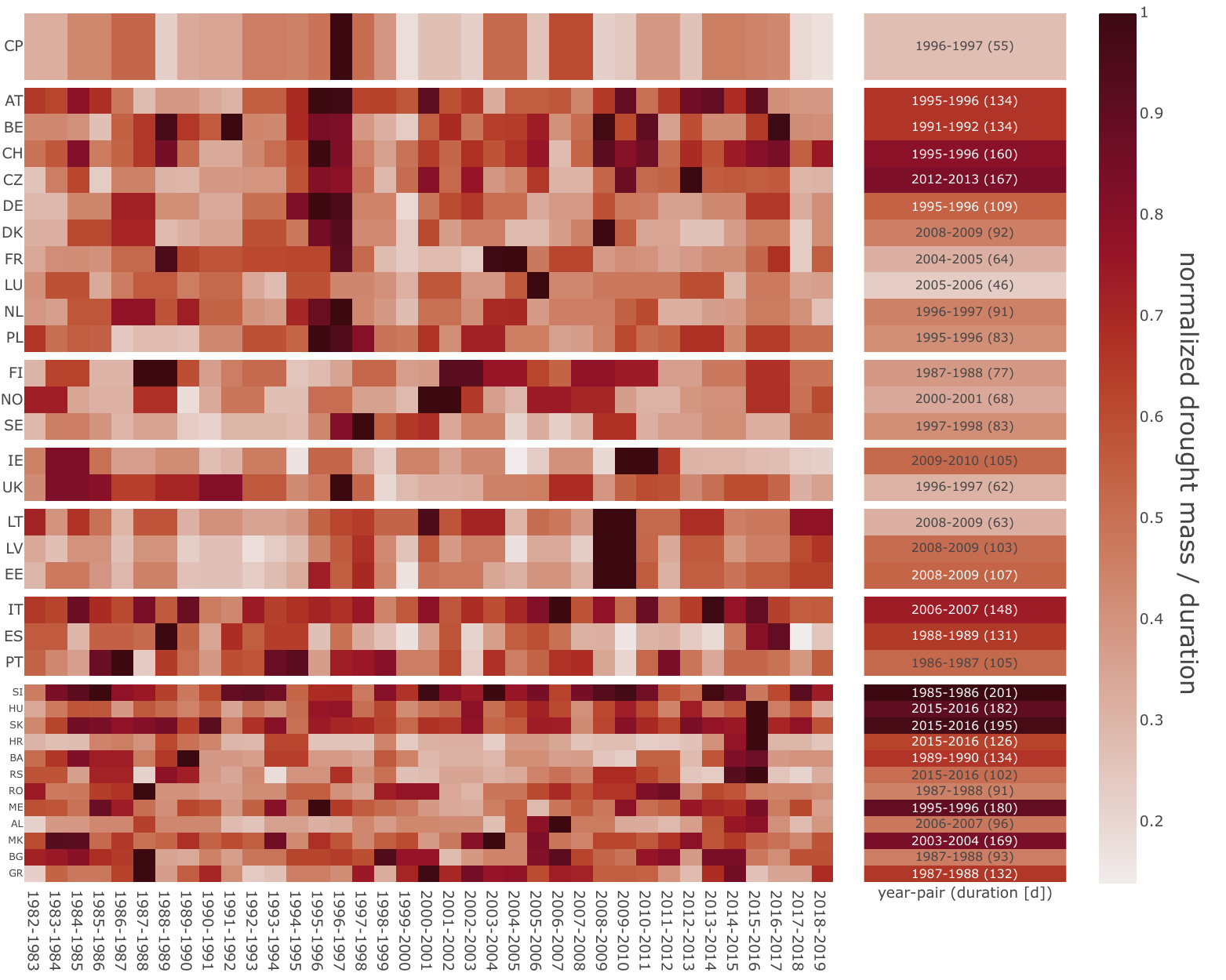}
\caption{Drought mass of identified most extreme winter drought events. \\ For each country or the European copperplate in the left panel, drought mass scores are normalized using the row-specific maximum. The colors of the right panel indicate the maximum duration of the event with the highest drought mass score normalized by the column-specific maximum, i.e.,~the maximum duration across all countries. The right panel also provides the year-pair with the most severe events as identified by the drought mass score per row and its corresponding duration in days.}
\label{fig:figure_8}
\end{figure}

The duration and severity of most extreme winter droughts captured by the drought mass metric varies substantially across years and countries (Figure~\ref{fig:figure_8}). Assuming perfect interconnection between countries, the most extreme event in the data occurred in the winter of 1996/97 and lasted 55~days. This European super drought was caused by pronounced and partially overlapping droughts in several Central European countries and the United Kingdom (Figure~\ref{fig:figure_8}). As the most pronounced events do not occur simultaneously in all countries, geographical balancing mitigates also such extreme droughts. Accordingly, the European super drought is substantially shorter than the most extreme droughts in nearly all isolated countries. Applying the drought mass metric to individual countries, we find the longest winter events in Eastern and Southern Europe. Further, smaller countries such as Slovenia (201~days, 1985/86) or Slovakia (195~days, 2015/16) tend to have longer extreme droughts than larger countries such as France (64~days, 2004/05), Sweden (83~days, 1997/98), Germany (109~days, 1995/96), or Spain (131~days, 1988/89). This is because smaller countries benefit less from geographical balancing within their borders.

\subsection{Geographical balancing inside the European renewable energy super drought}\label{sec:inside_drought}

\begin{figure}[htbp]
\centering
\noindent\includegraphics[width=\linewidth,keepaspectratio]{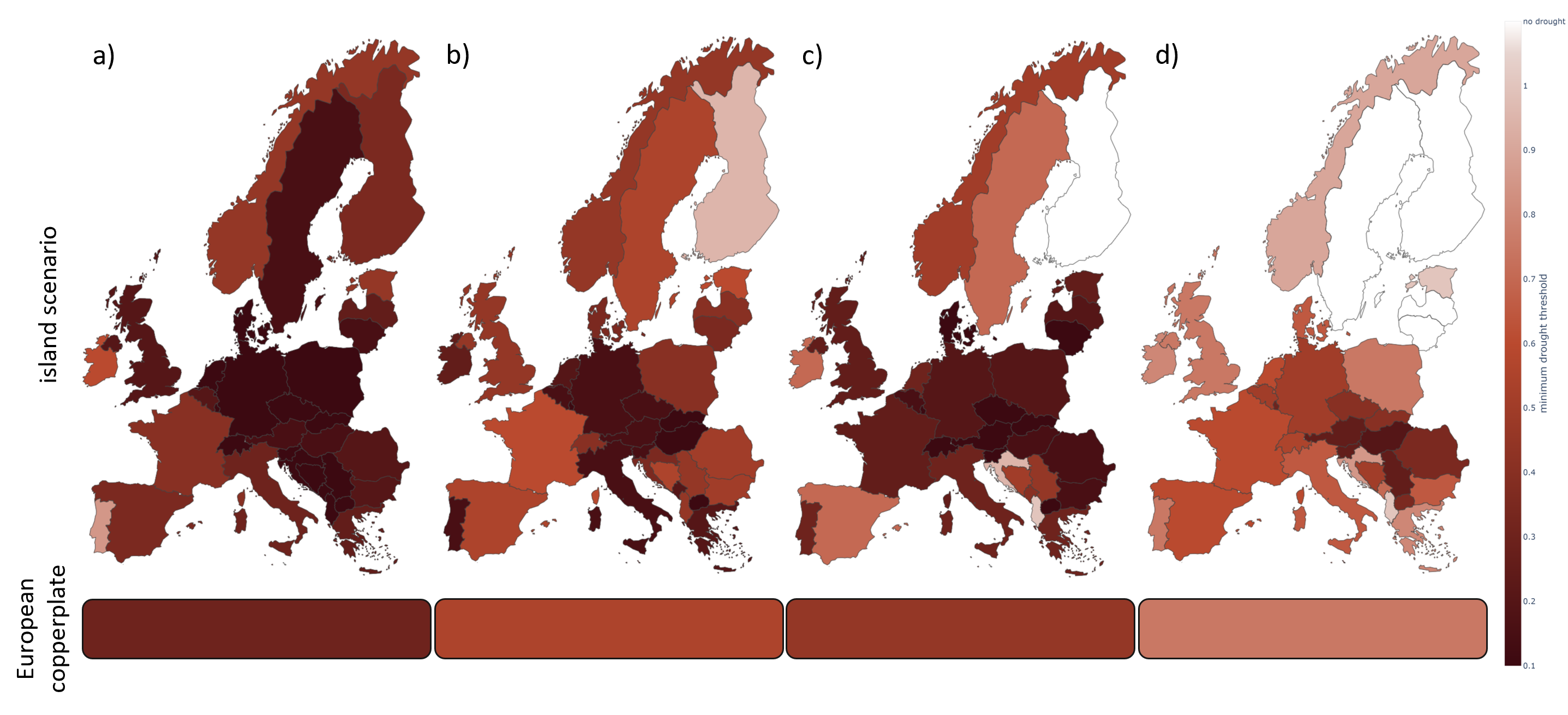}
\caption{Snapshots of drought severity during the European super drought in winter 1996/97. \\ Drought severity illustrated by the lowest applying threshold during the European super drought in winter 1996/97 for isolated countries (maps, top) and under the assumption of unconstrained geographical balancing for the European copperplate scenario (vertical bars, bottom).\\
a) Severe droughts covering almost all of Europe on 7~December 1996 at 8pm.\\
b) Severe droughts mainly in Central Europe on 19~December 1996 at 8am.\\
c) Severe droughts covering almost all of Continental Europe on 10~January 1997 at midnight.\\
d) Moderate droughts in most of Europe on 13~January 1997 at noon.}
\label{fig:figure_9}
\end{figure}

Figure~\ref{fig:figure_9} shows four snapshots of selected hours within the European portfolio super drought, as identified by the drought mass metric at the turn of the years 1996/97. Each panel shows the renewable drought severity present in the displayed hour for isolated countries in the island scenario (maps, top), illustrated by the lowest applying threshold. The corresponding drought severity in the European copperplate scenario, which is identical across all countries at each time step due to unconstrained geographical balancing, is illustrated as vertical bars at the bottom. While the super drought affects multiple European countries simultaneously, the severity varies across countries and over time. Comparing the island scenario and the European copperplate shows that renewable drought severity under unconstrained geographical balancing is always less pronounced than in the most affected isolated countries at the same hour. This is because the assumed European interconnection allows to mitigate energy shortages in one country by leveraging higher renewable availability in others. That is, geographical balancing is feasible even within the most severe pan-European drought event in the data. Further, even within the European super drought, there are hours with relatively high renewable availability in some countries (compare white-colored countries in Figures~\ref{fig:figure_9}c and~\ref{fig:figure_9}d). Supplementary Movie 1 provides a complementary animated graph, which shows the progression of the drought severity inside the European super drought, covering December 1996 and January 1997. The animation shows that the super drought comprises sequences of brief but more severe droughts, both in individual countries (different colors on maps on the left-hand side) and on the European level (different colors between maps on the right-hand side). It further shows that very severe droughts that cover most of Europe (e.g.,~also visible in Figure~\ref{fig:figure_9}a) are relatively brief and occur rarely, even inside the super drought event. That is, some renewable generation potential is always available somewhere in Europe. Even during the most extreme droughts, this can balance low renewable availability across technologies and regions to some extent.

\section{Discussion}\label{sec:discussion}


Variable renewable energy droughts manifest in a wide range of patterns, from brief and isolated to very long-lasting, contiguous events of varying severity. Our multi-threshold analysis reveals that drought frequency, return periods, and duration are highly sensitive to the chosen threshold. 
We thus argue that a multi-threshold approach is useful to adequately characterize heterogeneous and sequential patterns of renewable energy droughts of different severity. Single-threshold analyses, which are prevalent in the literature, have their merits in terms of simplicity. However, depending on the threshold chosen, they might fail to detect longer-lasting patterns and thus incompletely characterize extreme drought events. 

Based on a large data set and using a wide range of thresholds, we demonstrate that the complementarity of wind and solar power in Europe effectively reduces both the frequency and maximum duration of \ac{VRE} droughts. Analyzing isolated countries, we find that combining solar and wind power leverages a \textit{portfolio effect}, decreasing the maximum drought duration compared to single renewable technologies by on average 64\% (\ac{PV}), 52\% (onshore wind), or 47\% (offshore wind) in our European case study. This effect is particularly large for \ac{PV}, as wind power compensates for diurnal and seasonal shortages of solar energy. 
We further show that an unconstrained geographical balancing between European countries, which is unlikely to be achieved in practice, would give rise to a substantial \textit{balancing effect}, which in our case study shortens the longest \ac{PV}, onshore wind, offshore wind, or technology portfolio droughts by 1\%, 46\%, or 34\%, or 65\% on average. This effect is driven by an imperfect spatio-temporal correlation of extreme drought periods, especially concerning wind droughts.

To overcome the challenges of single-threshold analyses, we introduce a drought mass metric that enables an integrated assessment of the duration and severity of events. It reveals that extreme drought events of varying severity may emerge sequentially within contiguous low-availability periods lasting several weeks or months. These events affect multiple European countries simultaneously, yet with varying intensity. In most countries and in a perfectly interconnected Europe, the largest drought events occur in winter. In wind-dominated systems, however, the most severe events may occur during summer. We show that compound winter events, characterized by extreme technology portfolio droughts in combination with peak electricity demand, determine the major discharging period of long-duration electricity storage in a fully renewable European energy system.

The severity of such events varies substantially across years and countries, enabling geographical balancing to mitigate the need for system flexibility. Using the drought mass metric and drawing on renewable capacity assumptions from a policy-relevant scenario, we find that the most severe renewable technology portfolio drought observable in the data occurred in the winter of 1996/97. Even when assuming a perfectly interconnected European energy system, which would require massive transmission capacity expansion, this European super drought event would have lasted 55~days. The maximum droughts in individual, isolated countries are much longer, e.g., 109~days in Germany. This shows that even during the most pronounced European drought event, geographical balancing can be leveraged.

Previous research has highlighted the importance of diversifying variable renewable energy supply \cite{heide_seasonal_2010,jurasz_review_2020,harrison-atlas_temporal_2022} and advancing the integration of European power systems via transmission grid expansion \cite{schaber_parametric_2012,fursch_role_2013,roth_geographical_2023} for realizing cost-efficient renewable energy systems. Based on our analysis, we not only underscore this notion but extend it by arguing that technology portfolios and geographical balancing are also important strategies for effectively dealing with rare and extreme renewable energy droughts. We also show that the most severe \ac{VRE} portfolio droughts largely drive the need for long-duration storage in energy systems fully based on variable renewable energy sources. 

Our multi-threshold analysis facilitates a comprehensive characterization of renewable drought events in Europe, solely based on \ac{VRE} availability time series. Our approach, and particularly the drought mass metric, is not prone to artifacts that may arise in drought analyses relying on energy system modeling such as the duration of the planning horizon or technology cost assumptions. These may substantially affect optimal long-duration storage investments \cite{dowling_role_2020}. Yet, methodological challenges remain. Drought analyses purely based on renewable availability time series cannot take into account real-world interconnection levels, which are between our extreme assumptions in the island and copperplate scenarios. Drought characteristics of scenarios with cross-regional exchange are thus likely between these corner solutions. For instance, the duration of the European super drought is likely to last longer than the 55~days identified for the copperplate scenario, but shorter than identified here for isolated countries (Figure~\ref{fig:figure_8}), depending on future interconnection capacities between countries.

Further, the drought mass indicator does not consider a range of aspects that energy system models do. For instance, it does not capture periods with very high renewable availability that may arise between or after extreme droughts. Energy models with perfect foresight optimize long-duration storage operations considering not only droughts but also high-availability periods. This may result in major storage discharge periods that occur at different times or last longer than the most extreme drought periods identified by the drought mass indicator. 
In addition, the drought mass does not account for storage conversion losses or renewable curtailment. The latter may be required even within longer-lasting droughts in case of brief periods with very high availability that exceed storage charging capacities \cite{ruhnau_storage_2022}. It is possible to integrate conversion losses into drought identification methods, but this comes at the cost of higher complexity \cite{kittel_measuring_2024}. 

Due to computational limits, energy system models are often solved for a limited number of weather years, or even for only a single weather year. While substantial inter-annual variation in optimal energy model outcomes has been documented in the literature \cite{pfenninger_long-term_2016,staffell_using_2016,pfenninger_dealing_2017,zeyringer_designing_2018,collins_impacts_2018,schlott_impact_2018,kaspar_climatological_2019,hill_effects_2021,tong_geophysical_2021,grochowicz_intersecting_2023,gunn_spatial_2023,ruggles_planning_2024}, our analysis shows that extreme portfolio droughts also vary substantially across different years and regions. This likely aggravates inter-annual variations of energy model outcomes, particularly in terms of long-duration system flexibility needs. Accordingly, it appears desirable to corroborate the findings and conclusions of previous model-based energy system studies that are based on a small number of weather years or even only a single one and do not include extreme \ac{VRE} droughts. For example, the European \acl{TYNDP} 2022 draws on the single weather year 2009 \cite{entso-e_tyndp_2022}, while the newer \acl{TYNDP} 2024 uses the year 2019 \cite{entso-e_tyndp_2024}. Long-term energy scenarios developed for the European Commission generally lack transparency about actual weather years used, but appear to be based on a very limited number of weather years \cite{european_commission_directorate_general_for_energy_eu_2021}. The same holds true for influential climate neutrality scenarios for Germany \cite{bdi_klimapfade_2021,fraunhofer_isi_langfristszenarien_2024,prognos_klimaneutrales_2021,luderer_deutschland_2021}. Our analysis shows that the winter of 1996/97 including the pan-European super drought is particularly relevant for weather-resilient planning of the European energy system. It further indicates relevant weather years for single-country analysis. For any model-based energy system analysis that aims to shed light on how to deal with extreme renewable energy droughts, we suggest a two-step approach: first, use a large weather data set and identify the weather year with the most extreme drought event, purely drawing on renewable time series analysis as showcased in this paper; and second, ensure to include this particular year when running complex, numerically constrained energy system models.

Energy model planning horizons that are in line with a single calendar year further appear unsuitable for investigating weather-resilient future scenarios as extreme, storage-defining drought events may extend across the turn of years. Instead, we recommend adopting a planning horizon that captures relevant renewable seasonality patterns. For European settings, this may require at least bi-annual or single-year summer-to-summer planning horizons. Considering the typical annual cycles of European hydro reservoirs, which are not in the focus of our analysis, a spring-to-spring horizon might also be useful. Multi-annual planning horizons would be preferable but may come with increasing numerical challenges \cite{grochowicz_using_2024,ruggles_planning_2024}, and the length of the model period can impact optimal storage capacities \cite{dowling_role_2020}. 

Additionally, multi-sectoral and technology-rich energy system models with detailed grid representations often have to use aggregated temporal resolutions to reduce the computational burden \cite{nahmmacher_carpe_2016,kittel_temporal_2022,teichgraeber_time-series_2022,goke_adequacy_2022}. Our results indicate that the dynamics of storage-defining extreme drought events should adequately be reflected by representative time slices used in such models \cite{sanchez-perez_effect_2022}. This is even more relevant for integrated assessment models, which are also used for long-term energy system analyses \cite{gong_bidirectional_2023, schreyer_distinct_2024}, but usually come with even coarser time resolutions than energy system models. How to design respective temporal aggregation methods to properly represent extreme droughts that consist of a series of short extreme drought events during a prolonged period with low average renewable availability remains a methodological challenge.

Renewable drought analysis based on renewable availability time series, as conducted in this paper, is not intended for precisely quantifying long-duration storage requirements. Instead, it needs to be complemented by energy system modeling, which is required for evaluating flexibility needs in renewable energy systems for the reasons discussed above. Insights on extreme droughts can support the selection of relevant weather years for computationally intensive multi-carrier and multi-sectoral energy system models. Additionally, such insights can help explaining the outcomes of energy models, particularly concerning long-duration flexibility needs.

We see several promising avenues for future research. 
For example, a methodological comparison with alternative approaches that measure deviations from actual renewable availability from time-varying reference profiles \cite{stoop_climatological_2024, antonini_identification_2024}, would be of interest.
We note that a wide range of different datasets of renewable availability are used for energy system modeling, with sometimes conflicting model outcomes and policy recommendations \cite{kies_critical_2021}. Assessing the sensitivity of \ac{VRE} drought characteristics to such datasets \cite{kies_critical_2021,morin_reducing_2025} could provide valuable feedback to the meteorological research community that generates and continuously improves such datasets \cite{craig_overcoming_2022}. Further, the temporal extent of most data sets is limited to a few decades. Statistically robust findings for low-probability \ac{VRE} drought events with high impact are likely to require longer records of weather data, which also might involve a need for synthetic datasets, e.g.,~applied in Gangopadhyay et al. \cite{gangopadhyay_role_2022}. Moreover, drought analyses for modeled future weather data that incorporate the effects of global climate change would be desirable \cite{kapica_potential_2024,dolores_projections_2025}. Finally, using our methodology to characterize renewable energy droughts in other world regions and comparing these with our findings for Europe would be of interest. While parameters like the cut-off threshold of the drought mass metric ought to be reviewed, we argue that our multi-threshold framework is particularly useful for respective analyses in world regions where renewable availability is characterized by diurnal or seasonal variability, e.g.,~related to regional weather phenomena such as monsoon events.

\section{Methods}\label{sec:methods}

\subsection{Variable renewable energy drought definition and identification}\label{ssec:methods_vreda}

Our \ac{VRE} drought analysis is based on hourly \ac{VRE} availability factor time series. An hourly availability factor, also referred to as capacity factor, ranges between 0 and 1 and indicates the generation potential (in MWh) of a \ac{VRE} source in a specific region and hour normalized by its capacity (in MW). We use the \ac{VMBT} algorithm for drought identification as introduced and discussed in Kittel \& Schill \cite{kittel_measuring_2024}. This method searches for periods with a moving average below a given drought qualification threshold by iteratively decreasing the event duration. In each iteration, the algorithm sets the averaging interval to the respective event duration, starting with very long-lasting events and iteratively continuing to such that last only a few hours. A moving average below the drought threshold identifies a drought event. It is excluded from subsequent iterations, in which the averaging interval decreases further and additional (shorter) events are identified. 

This iterative procedure overcomes shortcomings of previous research \cite{kittel_measuring_2024}. It pools adjacent periods that independently may not qualify as \ac{VRE} drought, identifies unique events with the longest duration possible, avoids double counting as well as overlaps with adjacent events, and ensures that the full temporal extent of drought periods is captured \cite{kittel_measuring_2024}. To cover the full spectrum of \ac{VRE} droughts, the \ac{VMBT} algorithm iteratively searches for periods ranging in duration from two years to one hour in descending order. 

While threshold-based renewable drought identification methods are commonly applied in different strands of the academic literature, previous threshold-based analyses of renewable droughts across multiple regions, technologies, or technology portfolios often apply uniform, exogenously defined thresholds e.g.,~\cite{leahy_persistence_2013,cannon_using_2015,patlakas_low_2017,kaspar_climatological_2019,ohlendorf_frequency_2020,kies_critical_2021,potisomporn_extreme_2024,potisomporn_extreme_2024,abdelaziz_assessing_2024,mockert_meteorological_2023,mayer_probabilistic_2023,ohba_climatology_2022}). Such approaches implicitly assume that one unit of generation capacity has a uniform generation potential across all region-technology settings. However, annual generation potentials may differ substantially between regions and technologies due to varying meteorological conditions, which makes comparisons of the resulting drought characteristics difficult. To address this, threshold-based drought analyses should account for cross-regional and cross-technological variations in renewable generation potential when defining drought thresholds. Thresholds scaled relative to the global mean availability of each setting (termed here as ``relative thresholds''), reflect these variations \cite{kittel_measuring_2024}.

Additionally, as highlighted in the literature review, no consensus has yet been established regarding the selection of concrete threshold values. On the contrary, they often appear arbitrarily chosen or lack clear justification \cite{kittel_measuring_2024}.

To address both the challenge of comparability across regions and technologies and appropriate threshold value selection, we apply a wide range of relative drought thresholds, each scaled relative to the multi-year mean availability of the respective time series, i.e.,~each single technologies or different \ac{VRE} technologies or technology portfolios across countries. Threshold values $\tau_i$ range from 10\% to 100\% of the global mean availability factor over all investigated years, increasing in 5\% increments, i.e.,~$i \in [0.1, ..., 1.0]$. For instance, suppose $avail_{y,t,s}$ is the hourly availability factor in hour $t$ in year $y$ for the technology-region-setting $s$, and $T$ denotes the number all time steps across the entire set of investigated years (here $T =$ 38 years $\times t \in 8760$ hours per year $=332,880$ time steps), the relative threshold $\tau_i$ for the setting $s$ then relates to the value:

\begin{equation}
\tau_i = i \times \frac{\sum_{y,t} avail_{y,t,s}}{T}
\end{equation}

The broad range of relative thresholds $\tau_i$ with $i \in [0.1, ..., 1.0]$ allows us to capture a variety of renewable drought events. Very low thresholds capture potentially brief but very severe drought events with near-zero renewable availability (e.g., $\tau_{0.1}$), whereas high thresholds (e.g., $\tau_{1.0}$) are more likely to identify very long-lasting events or even full weather years with  below-average renewable availability. Figure~\ref{fig:figure_SI1} illustrates the relationship between these relative thresholds and absolute mean availability factors.

\subsection{Drought mass: a multi-threshold metric to identify storage-defining events}\label{ssec:method_drought_mass}

We introduce a drought mass metric to identify storage-defining drought events in a renewable energy systems. This metric identifies and ranks extreme events by integrating both drought duration and severity across a wide range of thresholds. As discussed in Section~\ref{ssec:max_duration}, very high relative thresholds close to $\tau_{1.0}$ are not meaningful, as they tend to capture very extended periods or even whole weather years with below-average renewable availability, instead of unusual drought events. For the drought mass metric computation, we therefore limit our analysis to drought events identified at thresholds $\tau_i$ with $i \in [0.1, 0.15, ..., 0.75]$, a set with cardinality $c = 14$. For each threshold $\tau_i$, we first generate binary time series for each year in the data, which assign the value~$1$ to hours that qualify as drought and $0$ otherwise. These time series are then paired into consecutive two-years periods (e.g.,~1982-1983, 1983-1984, 1984-1985, etc.), resulting in vectors with a magnitude of $T = 17,520$ hourly time steps per pair. This results in 37 year-pairs and accordingly allows analyzing 37 complete winter droughts. Note that the investigated data abstract from leap days. Next, we stack binary threshold-specific vectors to form a matrix of dimensions $T \times c$, where each row corresponds to a specific hour and each column to a threshold. Within each two-year period, we search for contiguous drought events identified at $\tau_{0.75}$, starting at hour $k$ and ending at hour $l$. For each of these intervals $[k, l]$, we compute the total number of drought hours across considered thresholds, i.e., we sum across all $c$ columns. The event with the highest cumulative multi-threshold score is then selected as the most extreme drought event for each weather year pair.

The drought mass metric equally weighs drought events identified across all considered thresholds. We chose a cut-off threshold $\tau_{0.75}$, which defines the event duration. Within an contiguous event, all thresholds $\tau_{i \leq 0.75}$ equally contribute to an event's overall score. We tested a wide range of alternative cut-offs and weighting schemes to identify the most effective drought mass design. The best alignment between events with the highest winter drought mass scores and major storage discharge events was achieved at a cut-off threshold of $\tau_{0.75}$ and a simple approach of equal weighting across all considered thresholds.

Countries with high shares of wind power in their capacity mix may experience the most extreme portfolio droughts in summer. Peak electricity demand periods often occur in winter, except in some South-European countries. To account for this, we compute a drought mass score for droughts occurring throughout the year and another one relating only to winter droughts excluding the period from May until September. When illustrating the relation of drought patterns and long-duration electricity storage use, we display both the most extreme summer and winter droughts if the highest drought mass score relates to summer droughts (compare regions with teal and purple boxes in Figure~\ref{fig:figure_7}). Conversely, if the highest drought mass score throughout the year relates to a winter drought, we mark only one event (compare regions with a gray box only in Figure~\ref{fig:figure_7}).

\subsection{Power sector modeling to determine long-duration storage use}\label{ssec:DIETER}

We use a stylized version of the open-source power sector model \ac{DIETER} to analyze the interaction between \ac{VRE} droughts and long-duration storage needs in a fully renewable European power sector. \ac{DIETER} is a linear optimization model that determines least-cost capacity and dispatch decisions based on an hourly resolution \cite{zerrahn2017,gaete-morales_dieterpy_2021}. Different versions of the model have been used to study various aspects of \ac{VRE} integration and their interaction with other flexibility options or sector coupling technologies \cite{zerrahn2018,schill_long-run_2018,schill_electricity_2020,schill_flexible_2020,kittel_renewable_2022,roth_geographical_2023,kirchem_power_2023,gong_bidirectional_2023,gaete-morales_power_2024, roth_power_2024}. Here, we use a model version that includes 33~European countries (EU27, the United Kingdom, Norway, Switzerland, and the Western Balkans). As very long-lasting \ac{VRE} droughts may span across the turn of a calendar year, we extend \ac{DIETER}'s planning horizon to two full years. 


The model features green hydrogen technologies, covering its generation, storage, and reconversion to electricity. We assume that hydrogen cavern storage can be expanded without restrictions in every country. This is a deliberate simplification, as we aim to illustrate the relation between \ac{VRE} portfolio droughts and long-duration electricity storage, but not to derive policy-relevant geographical allocations of hydrogen storage across Europe \cite{caglayan_technical_2020,talukdar_techno-economic_2024}. We focus on hydrogen used for long-duration electricity storage and abstract from additional hydrogen demand of other sectors, hydrogen imports from other world regions, and hydrogen exchange between countries.


\subsection{Input data}\label{ssec:data}

For the renewable drought analysis, we use country-level \ac{VRE} availability time series from the Pan-European Climate Database provided by the European Network of Transmission System Operators for Electricity \cite{de_felice_entso-e_2022}, comprising 38~weather years from 1982 to 2019. These data have been derived from reanalysis data, which has been converted to renewable availability by the Transmission System Operators. The database has been used for policy-relevant strategic reports, such as the \acf{TYNDP} 2022 or the European Resource Adequacy Assessment~2021. We use exogenous renewable capacity assumptions to generate the capacity-weighted \ac{VRE} portfolio time series as explained in the next paragraph. These assumptions stem from the \ac{TYNDP} 2022 (scenario Distributed Energy), which optimized these portfolios for a climate-neutral European power sector in 2050. For Germany, we update the capacity targets according to the latest government targets for 2045 \cite{bundesamt_fur_justiz_gesetz_2024}, in which Germany aims to achieve climate-neutrality.

For the power sector model, we additionally draw on policy-relevant renewable capacity mixes from the \ac{TYNDP} 2022 (scenario ``Distributed Energy''). Electricity demand profiles are derived from the European Resource Adequacy Assessment 2021 (target year 2025), including limited electrification of heat and transport. These profiles are scaled to the \ac{TYNDP} demand levels in 2050, and adjusted for net-importing and net-exporting countries to ensure they can be met by domestic renewable supply. Inter-annual variations are mainly driven by temperature differences. Other input data for the power sector model is provided, together with the model code, in the public repository \url{https://gitlab.com/diw-evu/projects/quantifying_the_dunkelflaute_energy_system_analysis}.

\subsection{Interconnection scenarios}\label{ssec:scenarios}

Multi-regional \ac{VRE} drought analyses based on \ac{VRE} availability time series can be conducted based on two extreme assumptions of electricity transmission between countries \cite{kittel_measuring_2024}: either perfect interconnection across all countries (``copper plate scenario'') or complete isolation of all countries (``island scenario''). For the former, all countries are treated as one single pan-European node. All regional time series are combined into a composite using capacity-weighted averages, with weights according to the capacity assumptions from \ac{TYNDP} 2022. For the latter scenario, thresholds are scaled as discussed above. Accordingly, regional \ac{VRE} portfolios are constructed using capacity-weighted averages of all contributing technologies. Note that more policy-relevant scenarios between these extreme cases, i.e., with limited interconnection capacity, cannot be applied to a purely time-series-based analysis without additional optimization or assumption-based heuristics. 

With the power sector model, we investigate three scenarios with varying degrees of interconnection between the 33~countries. First, we model the island scenario mentioned above without any cross-country exchange of electricity, i.e.,~domestic demand needs to be satisfied by domestic supply only. Second, we allow for unlimited exchange of electricity across countries, which represents the European copperplate scenario. These two scenarios mirror the ones used for renewable availability time series analysis discussed above. Third, we complement these extreme scenarios by allowing for a more policy-relevant cross-border exchange of electricity according to the Ten-Year-Network-Development-Plan 2022 (scenario ``Distributed Energy''). This additional scenario is used to illustrate the validity of the drought mass metric as a proxy for very severe \ac{VRE} droughts. Note that this third scenario with limited interconnection can only be investigated in that part of our analysis where the power sector model is used, as it leads to cost-minimizing flows of electricity between regions, but not in the main part that is purely based on renewable availability time series.

\section{Data availability}

The input data of the drought analysis tool are available in a public GitLab repository at \url{https://gitlab.com/diw-evu/variable_renewable_energy_droughts_analyzer}. The power sector model input data are provided in another public GitLab repository at \url{https://gitlab.com/diw-evu/projects/quantifying_the_dunkelflaute_energy_system_analysis}. High resolution, interactive, or animated versions versions of most figures presented in this article are available in a public Zenodo repository at \url{https://doi.org/10.5281/zenodo.13843390}.

\section{Code availability}

The code of the drought analysis tool is available in a public GitLab repository at \url{https://gitlab.com/diw-evu/variable_renewable_energy_droughts_analyzer}. The power sector model code is provided in another public GitLab repository at \url{https://gitlab.com/diw-evu/projects/quantifying_the_dunkelflaute_energy_system_analysis}.

\section*{Acknowledgments}
We thank the entire research group ``Transformation of the Energy Economy'' at the German Institute for Economic Research (DIW Berlin) for valuable inputs and discussions, as well as conference participants of the EGU General Assembly 2022 and 2025, the International Energy Workshop 2022, the Conference on Climate, Weather and Carbon Risk in Energy and Finance 2022, the Next Generation Energy and Climate Workshop 2022, the International Conference Energy \& Meteorology 2023, and the ENERDAY conference 2024 for valuable comments on earlier drafts. We acknowledge research grants by the Einstein Foundation (grant no.~A-2020-612) and by the German Federal Ministry of Education and Research via the ``Ariadne'' projects (Fkz 03SFK5NO \& 03SFK5NO-2).

\section*{Competing interests}

The authors declare no competing interests.

\section*{Author Contributions}\label{sec:author contributions}


\textbf{Martin Kittel}: Conceptualization (lead), methodology, software, investigation (equal), data curation, visualization, writing - original draft, review and editing (equal). \textbf{Wolf-Peter Schill}: Conceptualization (support), investigation (equal), writing - review and editing (equal), project administration, funding acquisition.

\newpage
\printbibliography


\newpage

\appendix

\renewcommand{\thesection}{SI}
\renewcommand{\thepage}{SI}
\renewcommand{\theequation}{\arabic{equation}}  

\global\long\def\thefigure{SI.\arabic{figure}}
\global\long\def\thetable{SI.\arabic{table}}
\global\long\def\thepage{\arabic{page}}  
\setcounter{figure}{0}
\setcounter{table}{0}
\setcounter{page}{1}
\setcounter{equation}{0}

\newcommand{\invisiblesection}[1]{%
  \phantomsection%
  \stepcounter{section}%
  \addcontentsline{toc}{section}{\protect\numberline{\thesection}#1}%
  }

\noindent\Large{\textbf{Multi-threshold time series analysis enables characterization of variable renewable energy droughts in Europe}}\\

\vspace{1cm}

\noindent\Large{\textbf{Supplementary information}}

\vspace{1cm}

\vspace{0.5cm}

\noindent\normalsize{Martin Kittel* and Wolf-Peter Schill}

\noindent\footnotesize{*Corresponding author: mkittel@diw.de}

\newpage





\section{Supplementary information}\label{sec:sup_inf}

\subsection{Supplementary Note 1: Employed thresholds}

\begin{figure}[H]
\centering
\subfloat[Solar \ac{PV}.\label{fig:figure_SI1a}]
    {{\includegraphics[width=.47\textwidth]{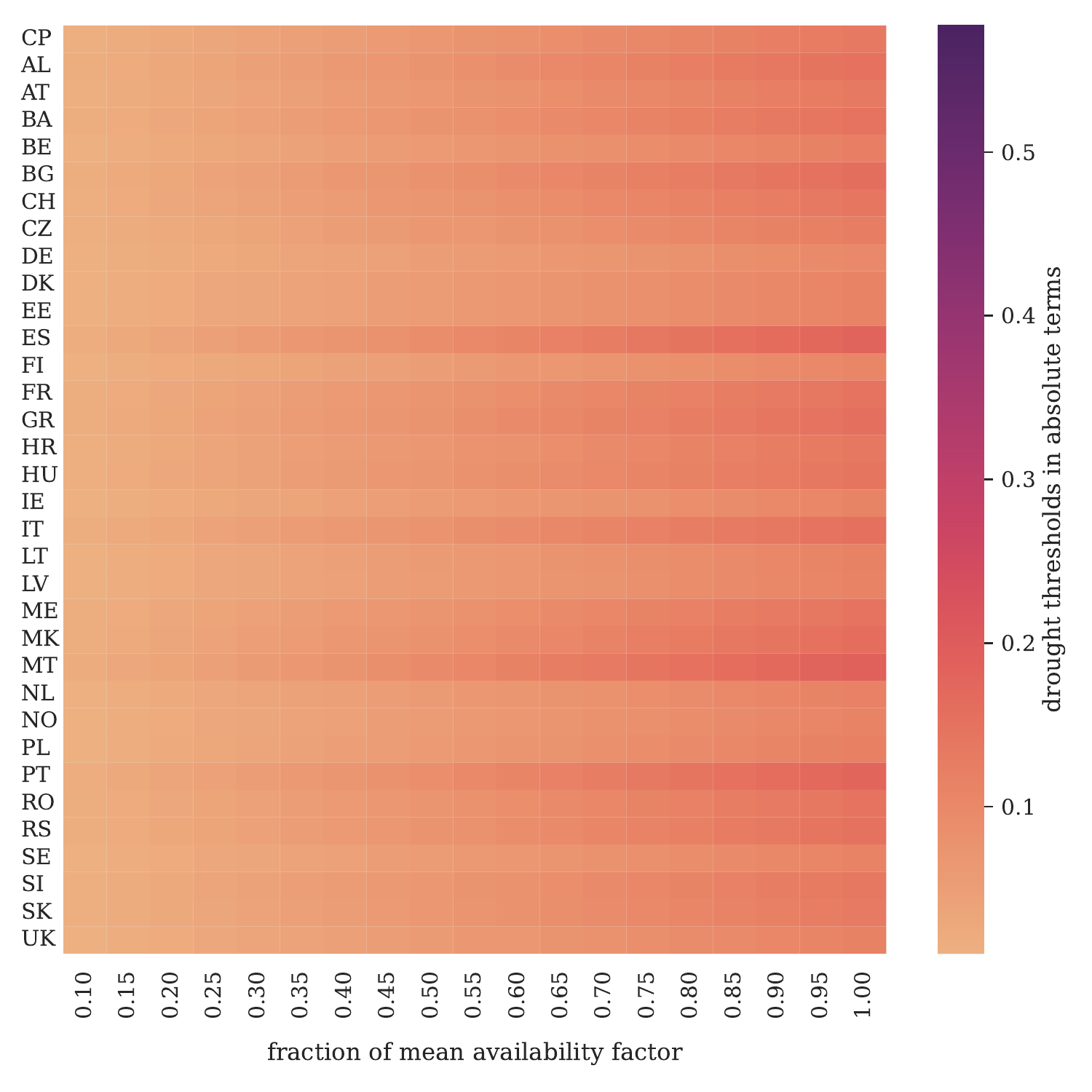}}}
\subfloat[Onshore wind.\label{fig:figure_SI1b}]
    {{\includegraphics[width=.47\textwidth]{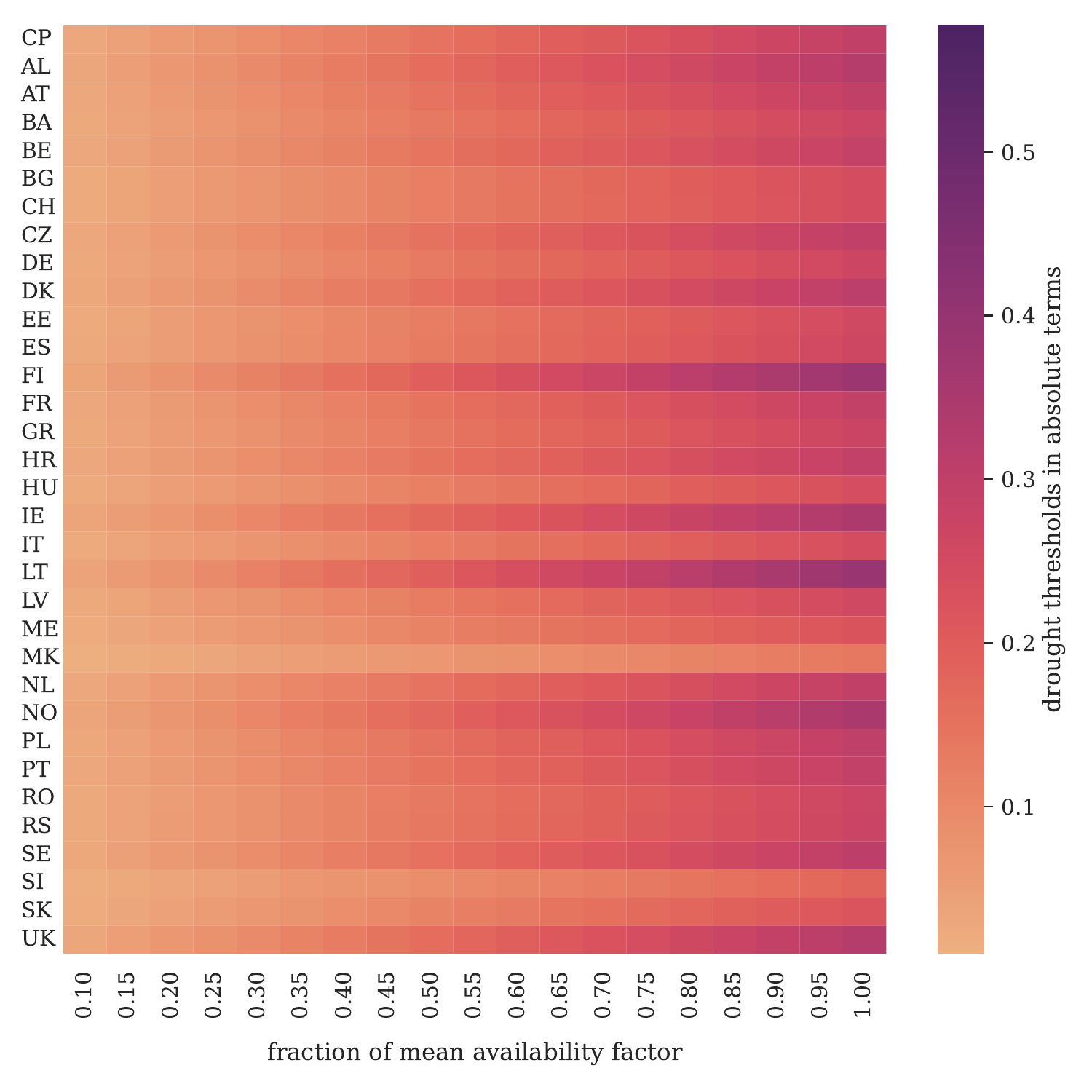}}}
\quad
\subfloat[Offshore wind.\label{fig:figure_SI1c}]
    {{\includegraphics[width=.47\textwidth]{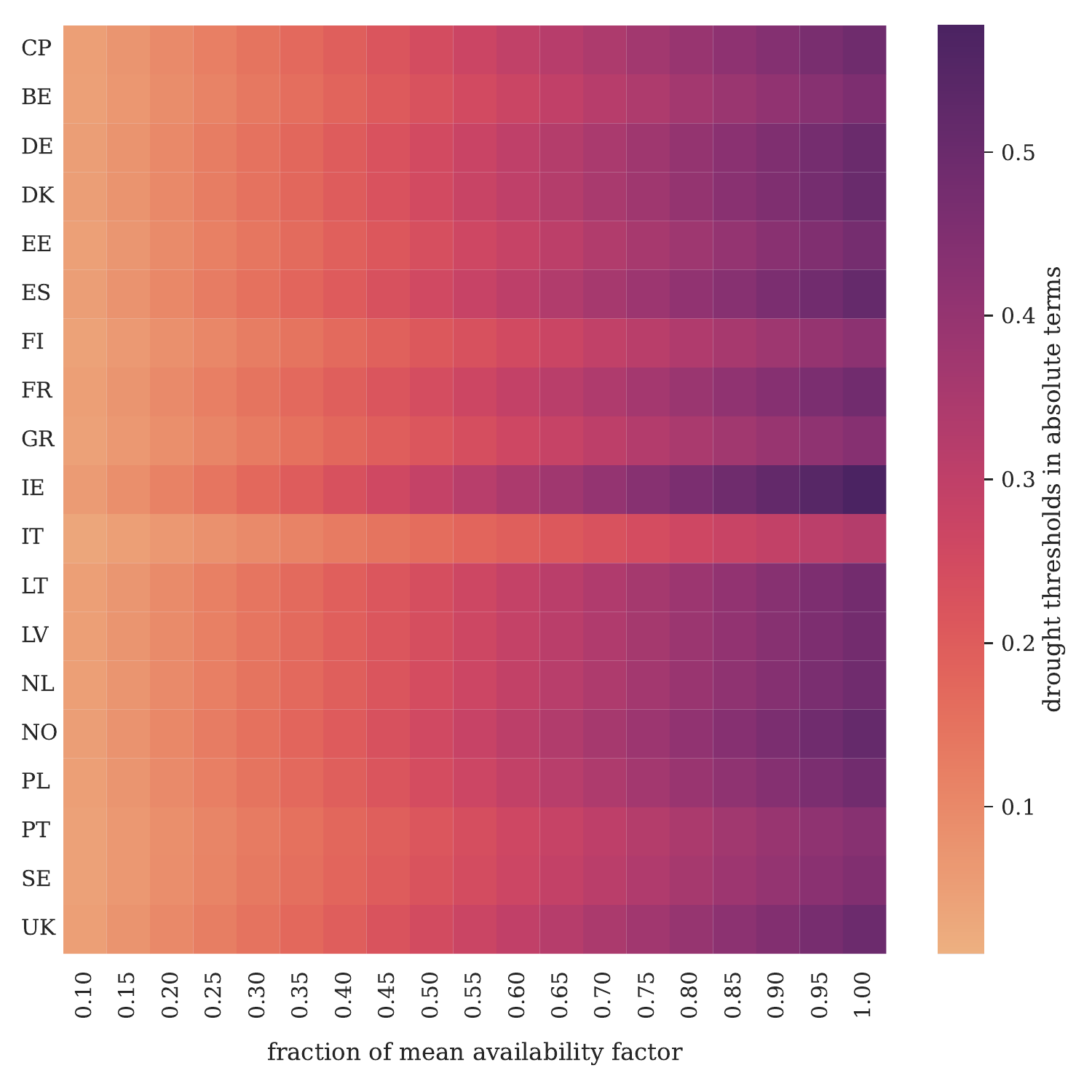}}}
\quad
\subfloat[Renewable technology portfolio.\label{fig:figure_SI1d}]
    {{\includegraphics[width=.47\textwidth]{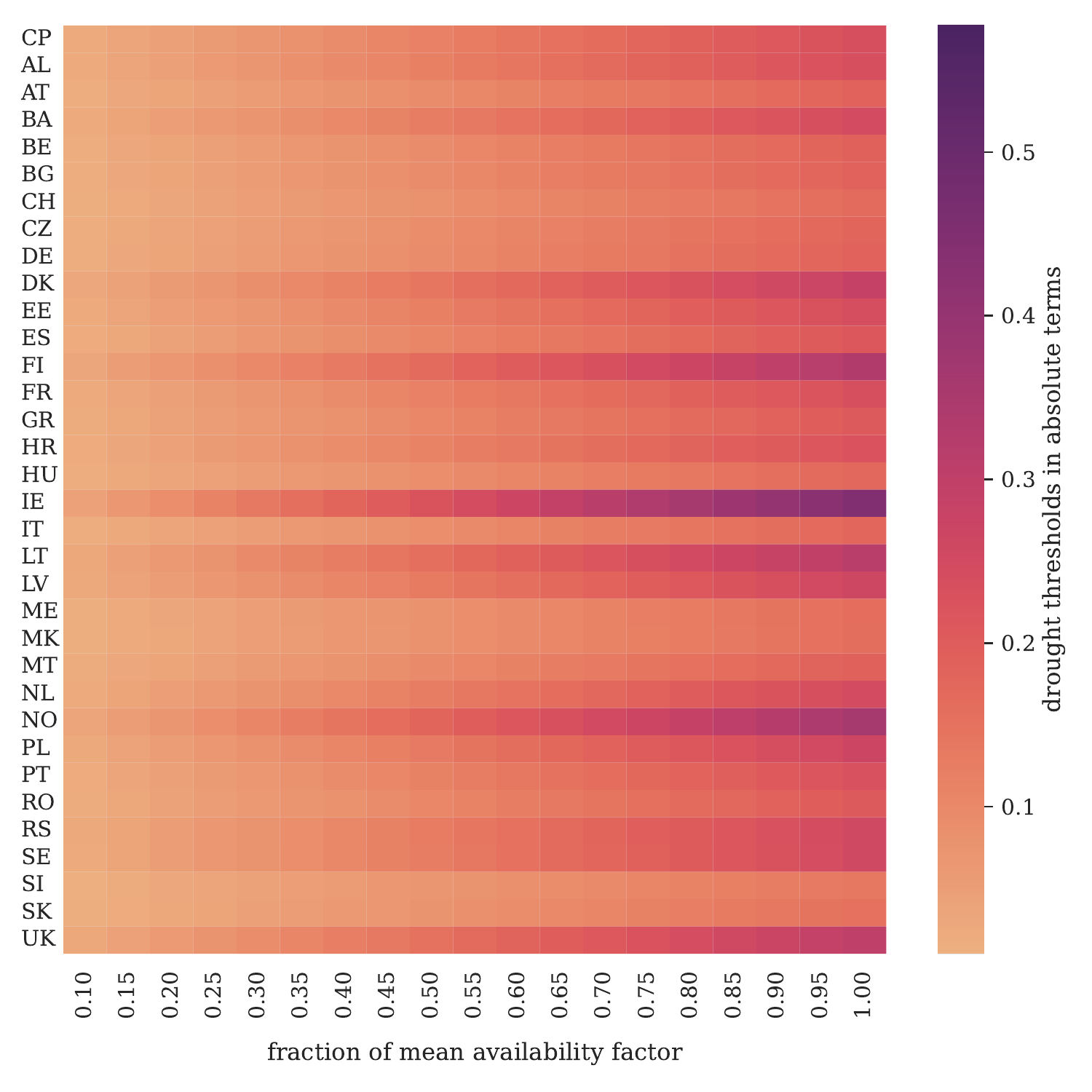}}}\\
\caption{Relative thresholds used for drought identification in absolute terms. \\ The renewable technology portfolio thresholds are based on capacity-weighted composite time series.}%
\label{fig:figure_SI1}%
\end{figure}

Figure~\ref{fig:figure_SI1} illustrates how the relative thresholds used in the search algorithm relate to absolute availability factors for each \ac{VRE} technology and for the renewable technology portfolio. Absolute thresholds are generally lower for solar \ac{PV} than for wind power because of the lower average availability of solar \ac{PV}.

\subsection{Supplementary Note 2: Additional illustrations of identified drought patterns}\label{ssec:sup_inf_drought_patterns}

\begin{figure}[H]
\centering
\noindent\includegraphics[width=\linewidth,keepaspectratio]{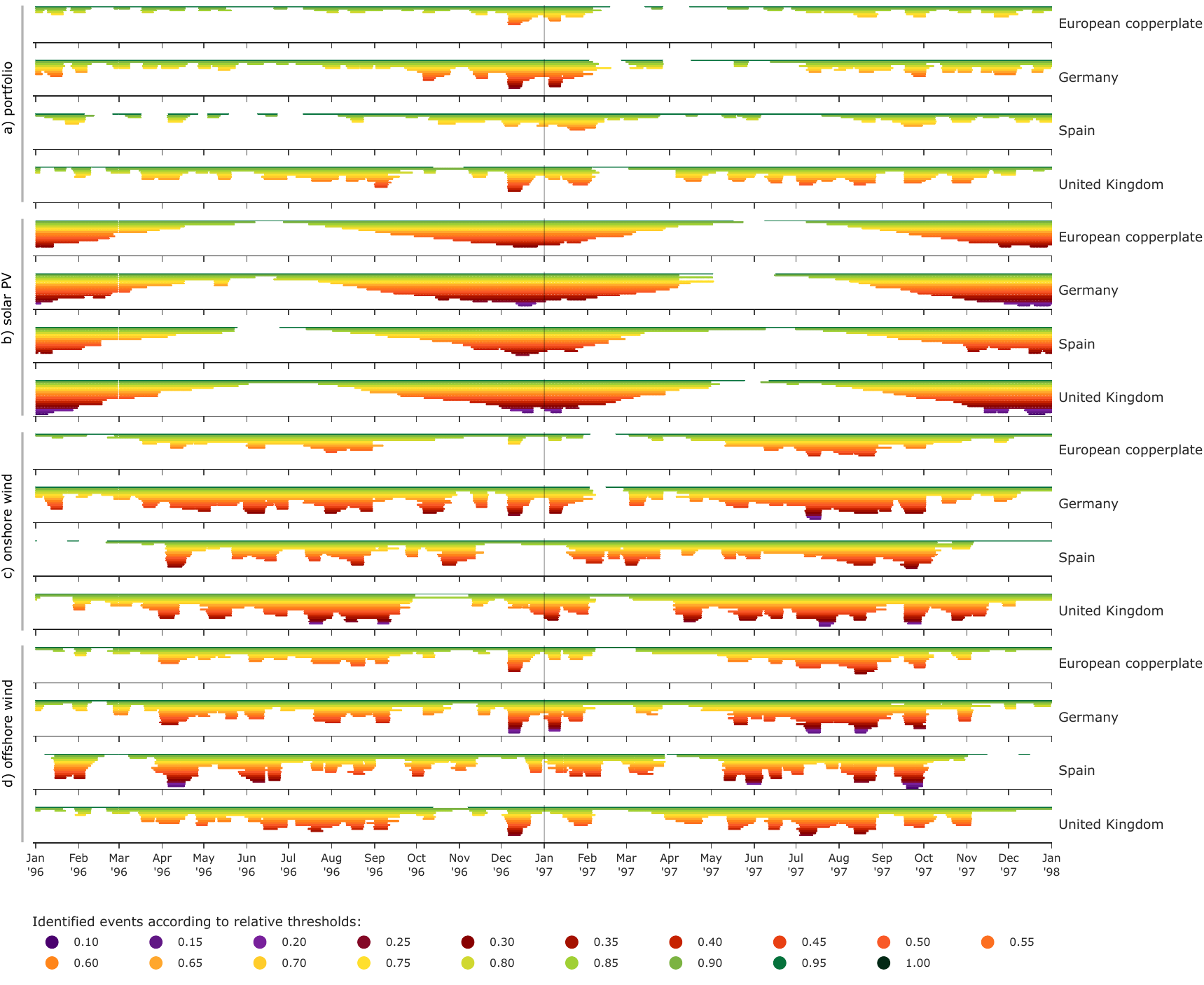}
\caption{Identified drought patterns in 1996 and 1997 across all employed thresholds. \\ The dates shown on the bottom axis correspond to all technology-region-specific panels. For each panel, the stacked horizontal bands indicate drought occurrences for the color-coded threshold of a particular region. To illustrate very persistent shortage situations, only droughts lasting longer than one week are displayed.}
\label{fig:figure_SI2}
\end{figure}

\begin{figure}[H]
\centering
\noindent\includegraphics[width=\linewidth,keepaspectratio]{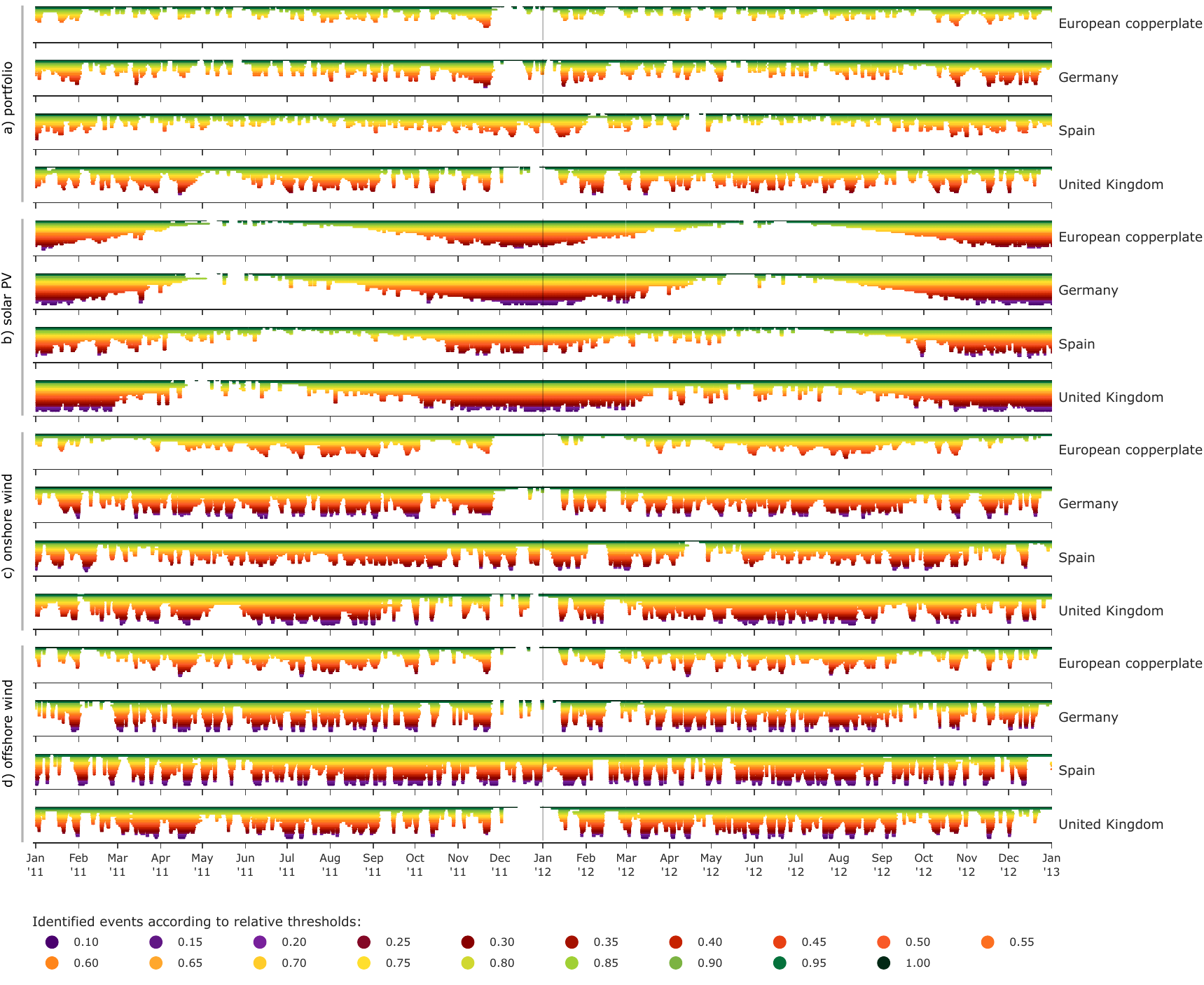}
\caption{Identified drought patterns in 2011 and 2012 across all employed thresholds. \\ The dates shown on the bottom axis correspond to all technology-region-specific panels. For each panel, the stacked horizontal bands indicate drought occurrences for the color-coded threshold of a particular region. To illustrate persistent patterns, only droughts lasting longer than one day are displayed. In Spain, low \ac{PV} seasonality is accompanied by wind droughts, translating into portfolio droughts arising across the turn of the year 2011/12. In contrast, due to the exceptionally high wind availability in North Europe, no portfolio droughts occur across the turn of the year in these countries.}
\label{fig:figure_SI3}
\end{figure}

\begin{figure}[H]
\centering
\noindent\includegraphics[width=\linewidth,keepaspectratio]{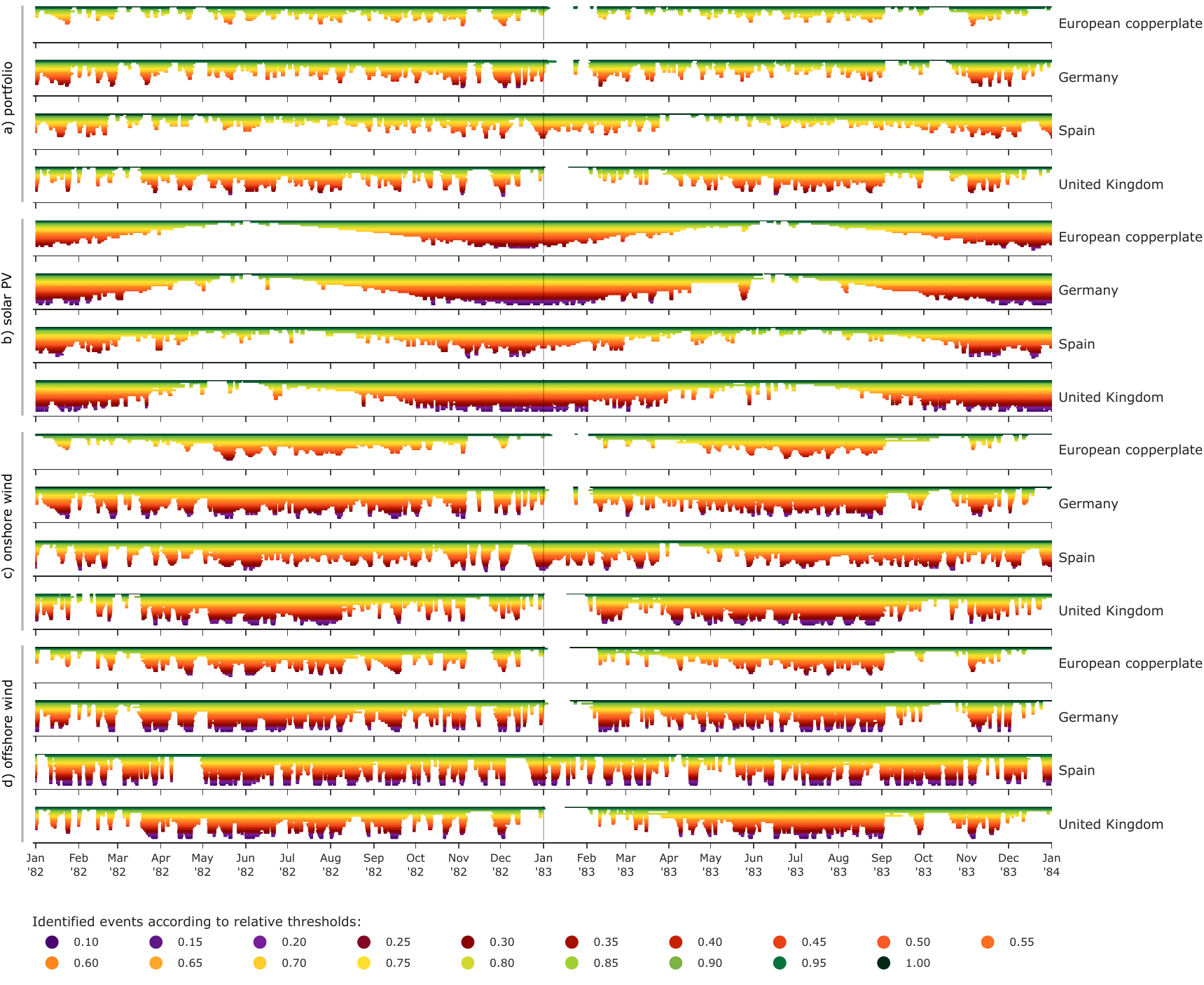}
\caption{Identified drought patterns in 1982 and 1983 across all employed thresholds. \\ The dates shown on the bottom axis correspond to all technology-region-specific panels. For each panel, the stacked horizontal bands indicate drought occurrences for the color-coded threshold of a particular region. To illustrate persistent patterns, only droughts lasting longer than one day are displayed. Except for Spain, no portfolio droughts occur at the beginning of a calendar year due to the absence of wind droughts in Northern Europe.}
\label{fig:figure_SI4}
\end{figure}

\begin{figure}[H]
\centering
\noindent\includegraphics[width=\linewidth,keepaspectratio]{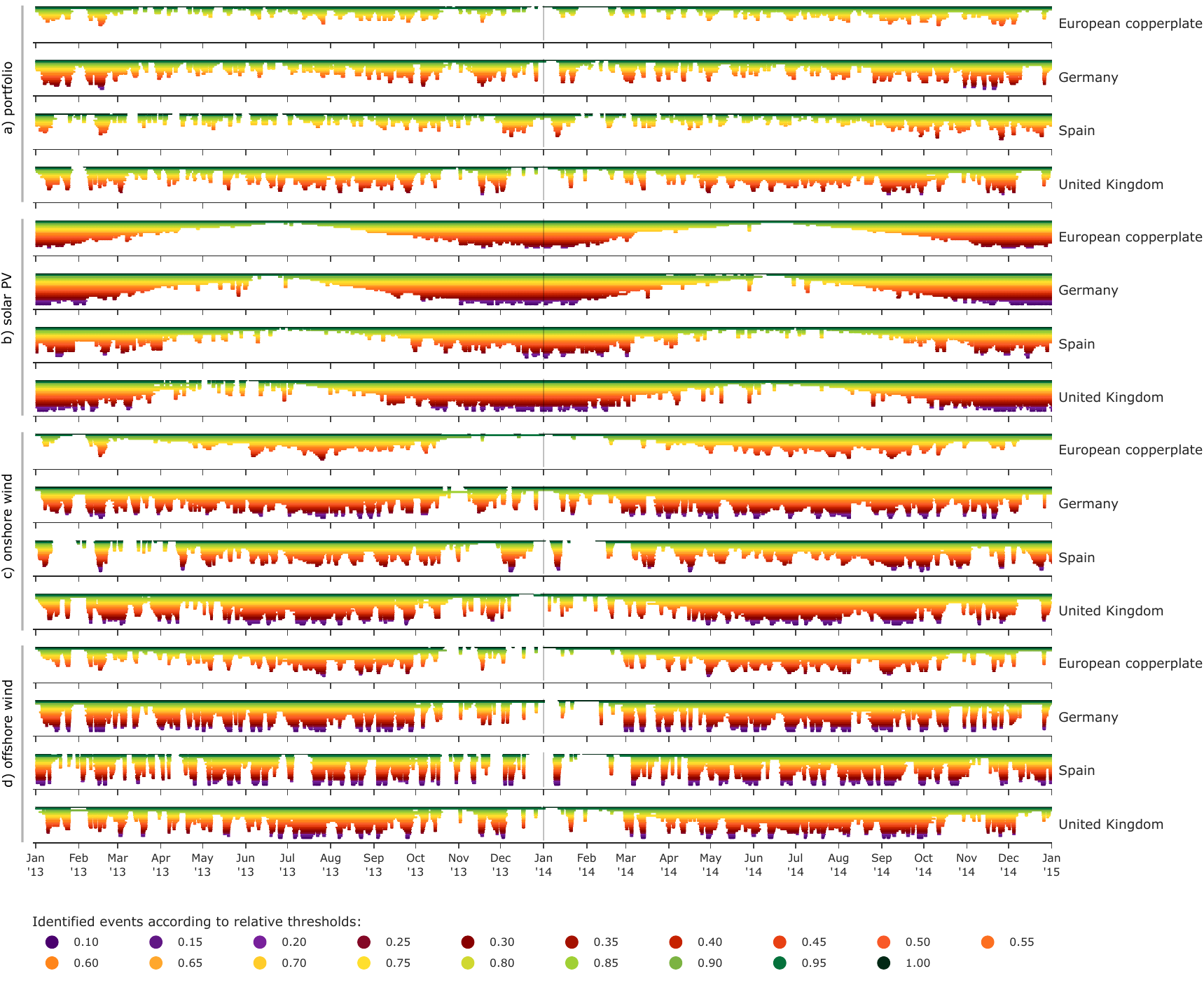}
\caption{Identified drought patterns in 2013 and 2014 across all employed thresholds. \\ The dates shown on the bottom axis correspond to all technology-region-specific panels. For each panel, the stacked horizontal bands indicate drought occurrences for the color-coded threshold of a particular region. To illustrate persistent patterns, only droughts lasting longer than one day are displayed. Brief wind droughts are infrequent in winter but longer-lasting, more severe, and more frequent in summer. Therefore, less severe portfolio droughts occur in the winter of 2013/14.}
\label{fig:figure_SI5}
\end{figure}

Figure~\ref{fig:figure_SI2} illustrates identified droughts lasting longer than one week or more, filtering out briefer events. This highlights very long-lasting below-average wind periods that may encompass multiple drought events of varying severity and solar seasonality. Combined in a portfolio, these periods are less severe (portfolio effect) and are further mitigated when assuming unconstrained geographical balancing (balancing effect). Figures~\ref{fig:figure_SI3},~\ref{fig:figure_SI4}, and~\ref{fig:figure_SI5} show identified drought patterns lasting at least one day for the years 2011/12, 1982/83, and 2013/14. 

\begin{figure}[htbp]
\centering
\noindent\includegraphics[width=\linewidth, keepaspectratio]{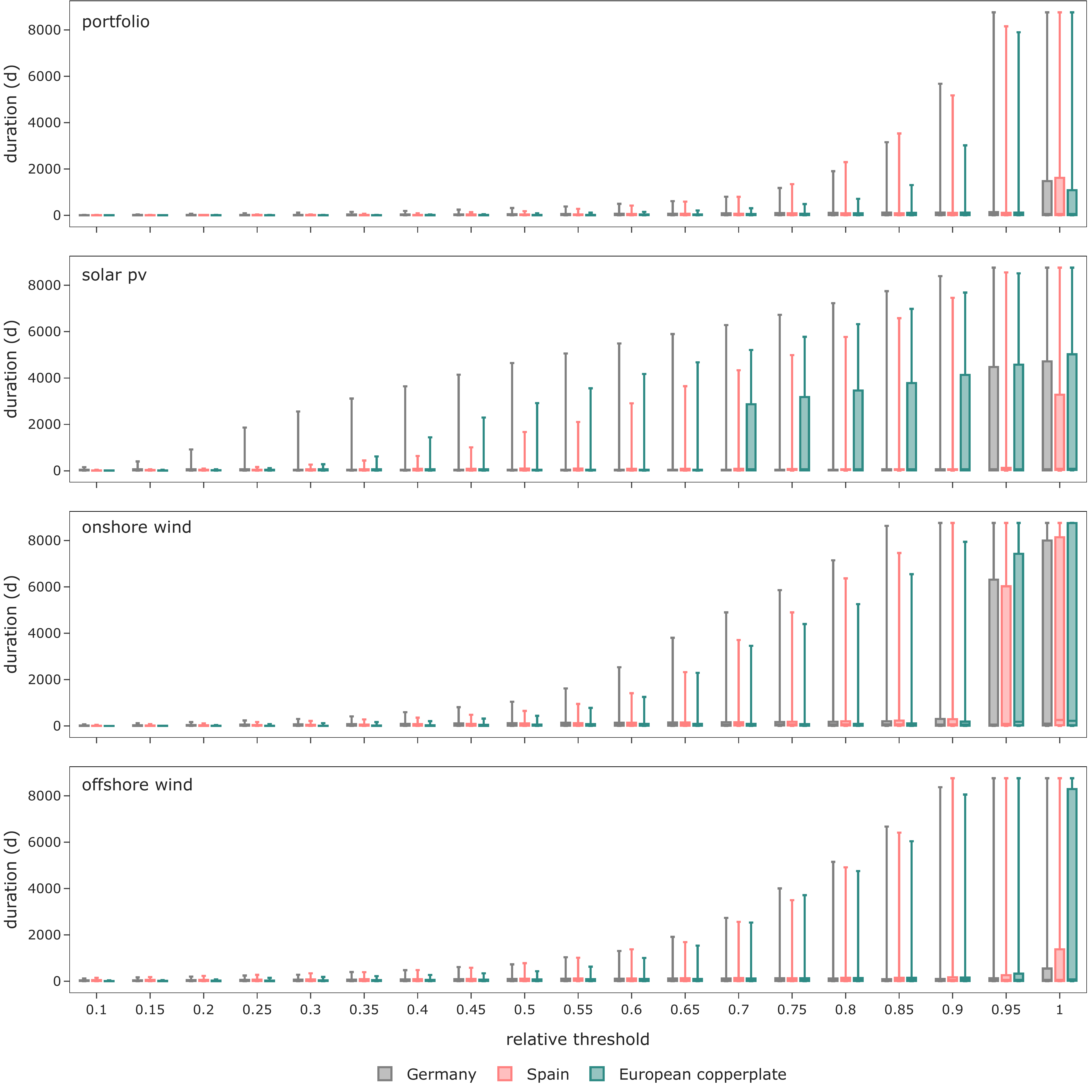}
\caption{Drought duration distribution for selected regions. \\ Box plot of drought duration distribution across all investigated relative drought thresholds $\tau_i$ with $i \in [0.1,..., 1]$. For illustration, the box plot whiskers relate to the 1 and 99 percentile of the distribution.}
\label{fig:figure_SI6}
\end{figure}

Figure~\ref{fig:figure_SI6} illustrates the threshold-specific distribution of all identified droughts in the data across all weather years for Germany and Spain, as examples of central and southern European countries with different wind and solar energy resources, and for the European copperplate scenario. The distributions confirm the \textit{portfolio effect} from combining complementary wind and solar \ac{PV} profiles into a technology portfolio discussed above, both in terms of mean and maximum duration (compare the panels of single \ac{VRE} technologies to the portfolio one in Figure~\ref{fig:figure_SI6}).

\newpage

\subsection{Supplementary Note 3: Additional illustrations of frequency-duration distributions}\label{ssec:sup_inf_frequency} 

\begin{figure}[H]
\centering
\noindent\includegraphics[width=\linewidth,height=\textheight, keepaspectratio]{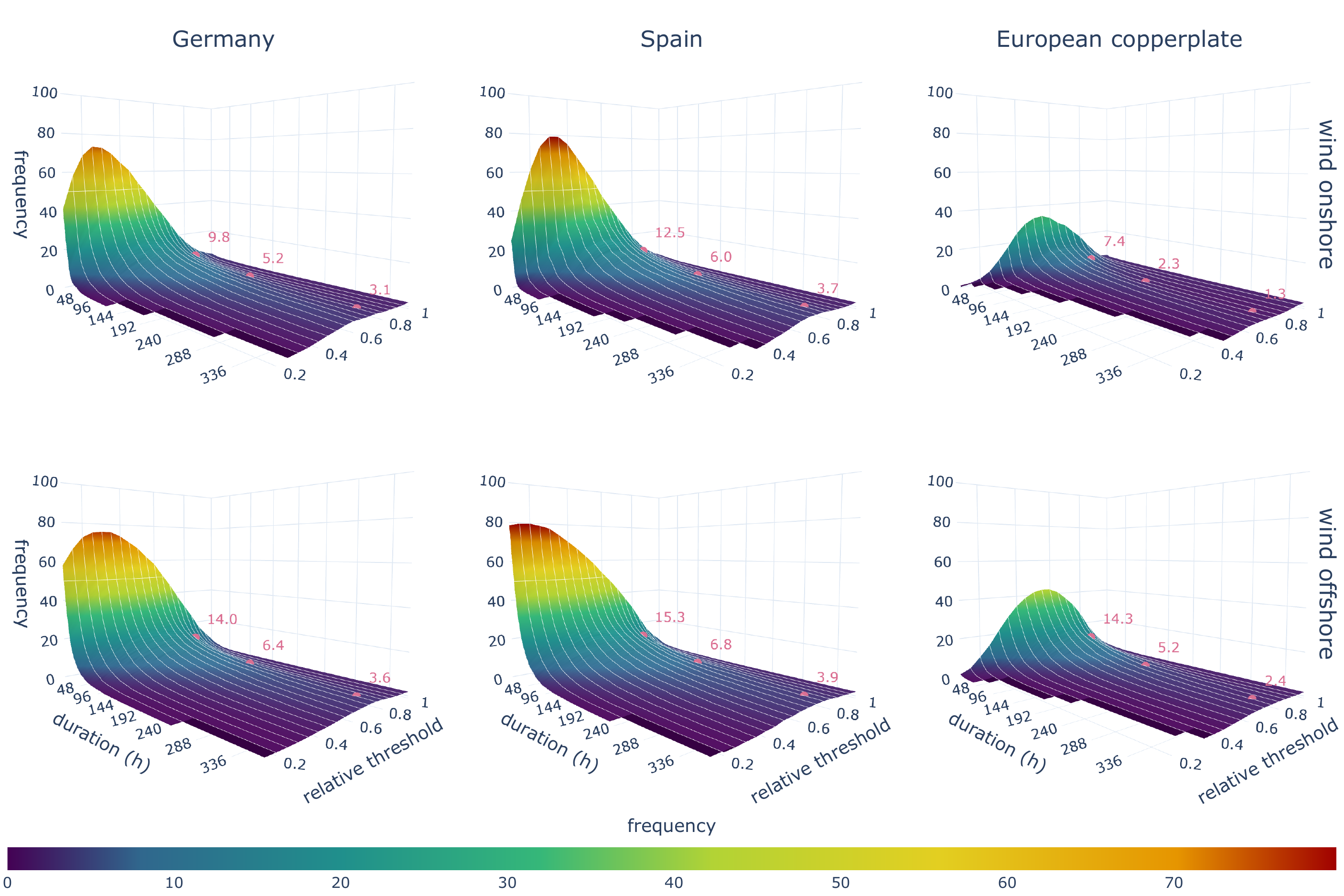}
\caption{Example of frequency-duration distributions of drought events. \\ Cumulative frequency-duration distributions of wind drought events across all investigated relative drought thresholds $\tau_i$ with $i \in [0.1,..., 1]$, sorting the yearly frequencies of all events that are at least as long as a given duration. White space indicates the absence of droughts for given thresholds in the data. The contour lines represent the threshold-specific yearly frequency. For illustration, the distributions are truncated at 360 hours, i.e., they show events with a maximum duration of just above two weeks. Yearly frequencies of events lasting at least two days, one week, and a fortnight are marked for a relative threshold of $\tau_{0.75}$.}
\label{fig:figure_SI7}
\end{figure}

\begin{figure}[H]
\centering
\noindent\includegraphics[width=\linewidth,height=\textheight, keepaspectratio]{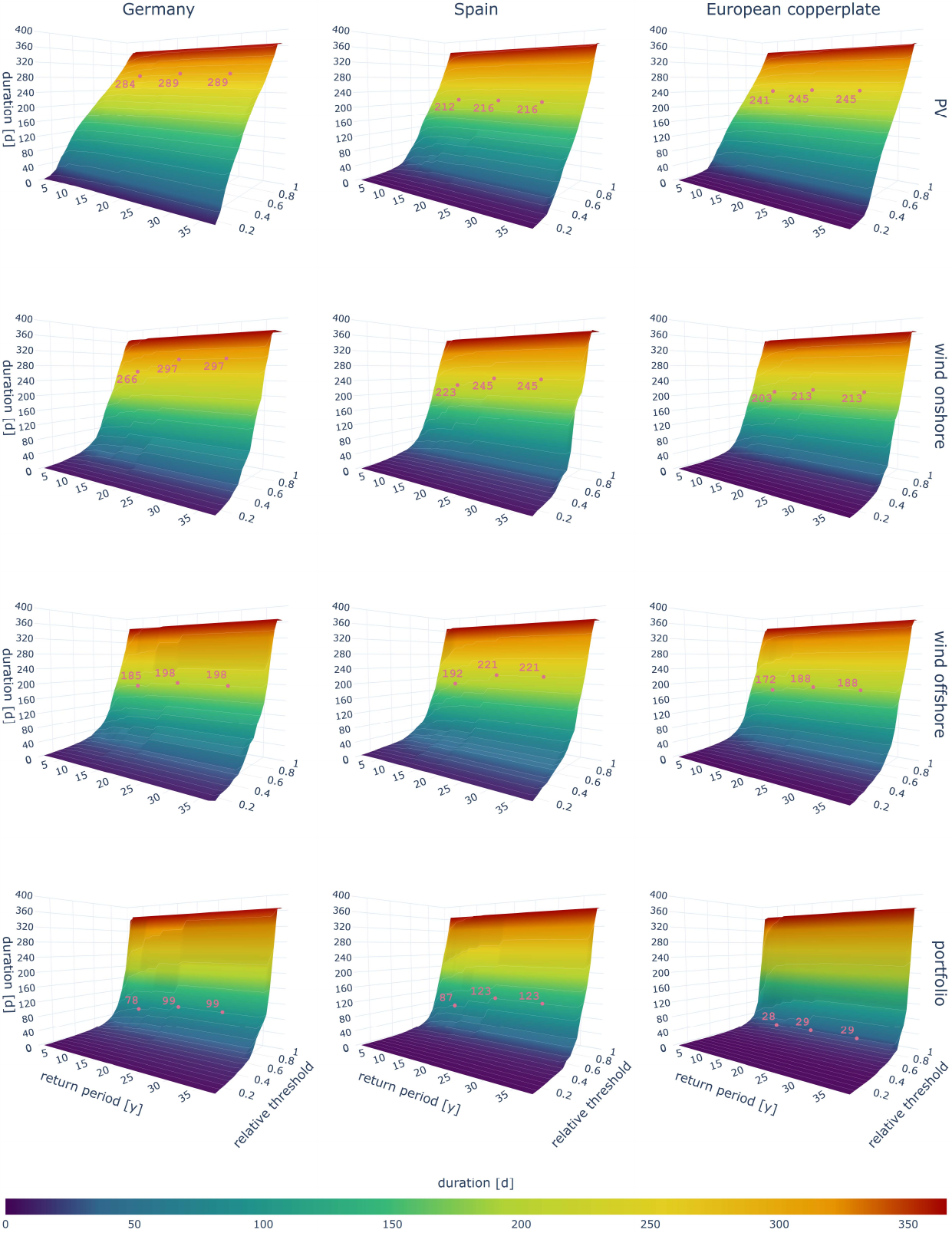}
\caption{Example of frequency-duration distributions of drought events that may last up to one full year. \\ Cumulative frequency-duration distributions of drought events across all investigated thresholds $\tau_i$ with $i \in [0.1, ..., 1]$ , sorting the frequencies of all events that are at least as long as a given duration. White space indicates the absence of droughts for given thresholds in the data. The contour lines represent the threshold-specific frequency. An interactive version of this Figure which allows zooming and rotating is available online \cite{kittel_high_2025}.}
\label{fig:figure_SI8}
\end{figure}

\begin{figure}[H]
\centering
\noindent\includegraphics[width=\linewidth, keepaspectratio]{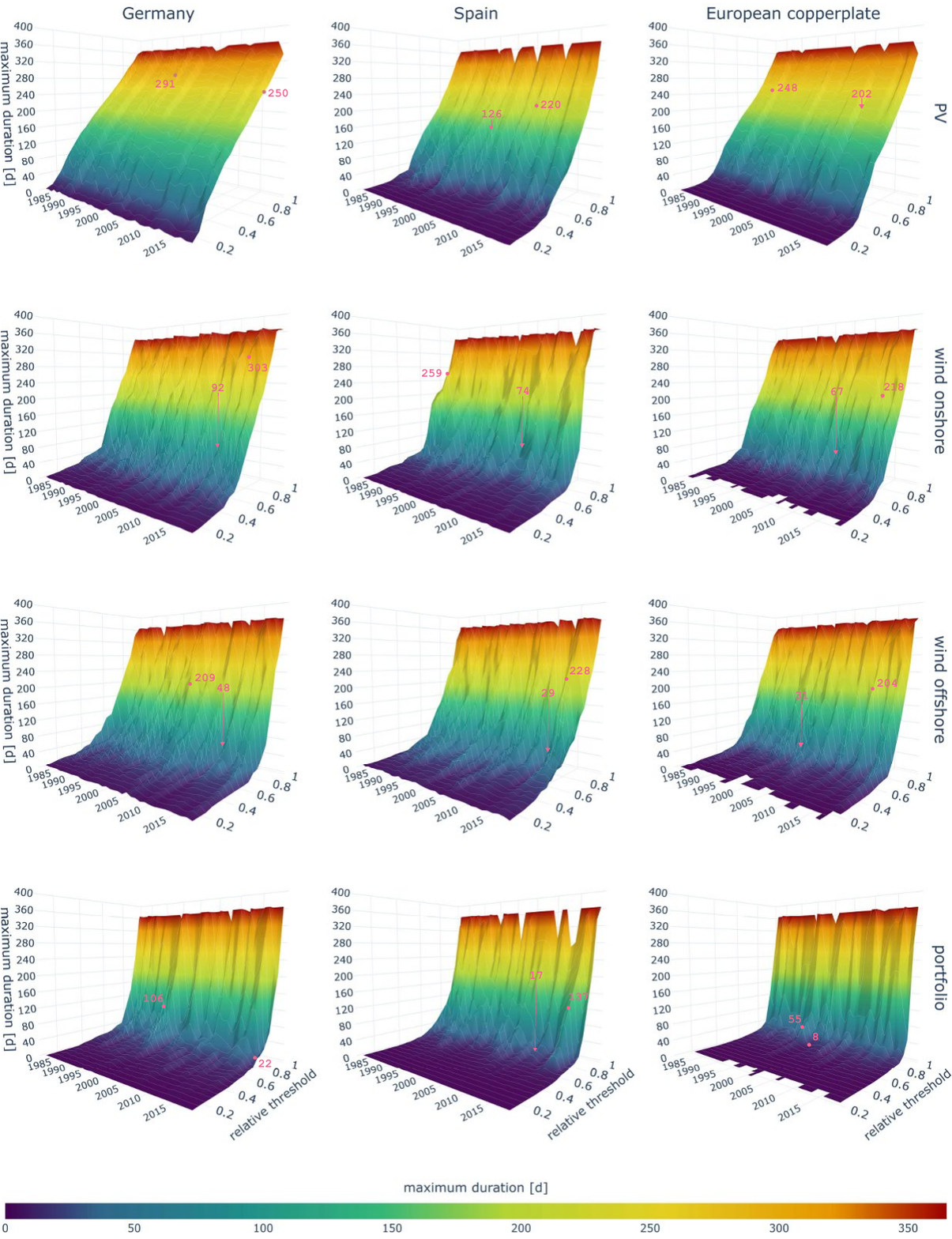}
\caption{Example of frequency-duration distributions of drought events under the assumption of unconstrained geographical balancing. \\ Cumulative frequency-duration distributions of drought events across all investigated thresholds $\tau_i$ with $i \in [0.1, ..., 1]$, sorting the frequencies of all events that are at least as long as a given duration. White space indicates the absence of droughts for given thresholds in the data. The contour lines represent the threshold-specific frequency. For illustration, the distributions are truncated at 360 hours, i.e., they show events with a maximum duration of just above two weeks. Frequencies of events lasting at least two days, one week, and a fortnight are marked for a relative threshold $\tau_{0.75}$. An interactive version of this Figure which allows zooming and rotating is available online \cite{kittel_high_2025}.}
\label{fig:figure_SI9}
\end{figure}

In contrast to Figure~\ref{fig:figure_2}, Figure~\ref{fig:figure_SI7} shows the cumulative frequency-duration distributions for on- and offshore wind droughts for Germany, Spain, and the European copperplate scenario with adjusted frequency-axis range. The \textit{balancing effect} mitigates the frequency and duration of wind droughts, particularly those found by lower thresholds. Figure~\ref{fig:figure_SI8} illustrates the cumulative frequency-duration distributions for droughts lasting up to a full year for Germany, Spain, and the European copperplate scenario. Only higher thresholds identify droughts lasting longer than a few weeks. For thresholds $\tau_{i< 1.0}$, portfolio droughts are generally briefer than single-technology droughts. To raise complementary insights on seasonality, Figure~\ref{fig:figure_SI9} shows seasonally differentiated distributions for the copperplate scenario. In general, longer droughts are more frequent in winter than in summer. Brief \ac{PV} droughts are more frequent in summer and are detected by higher thresholds.

\subsection{Supplementary Note 4: Additional illustrations of return periods}\label{ssec:sup_inf_return_period} 

\begin{figure}[htbp]
\centering
\noindent\includegraphics[width=\linewidth, keepaspectratio]{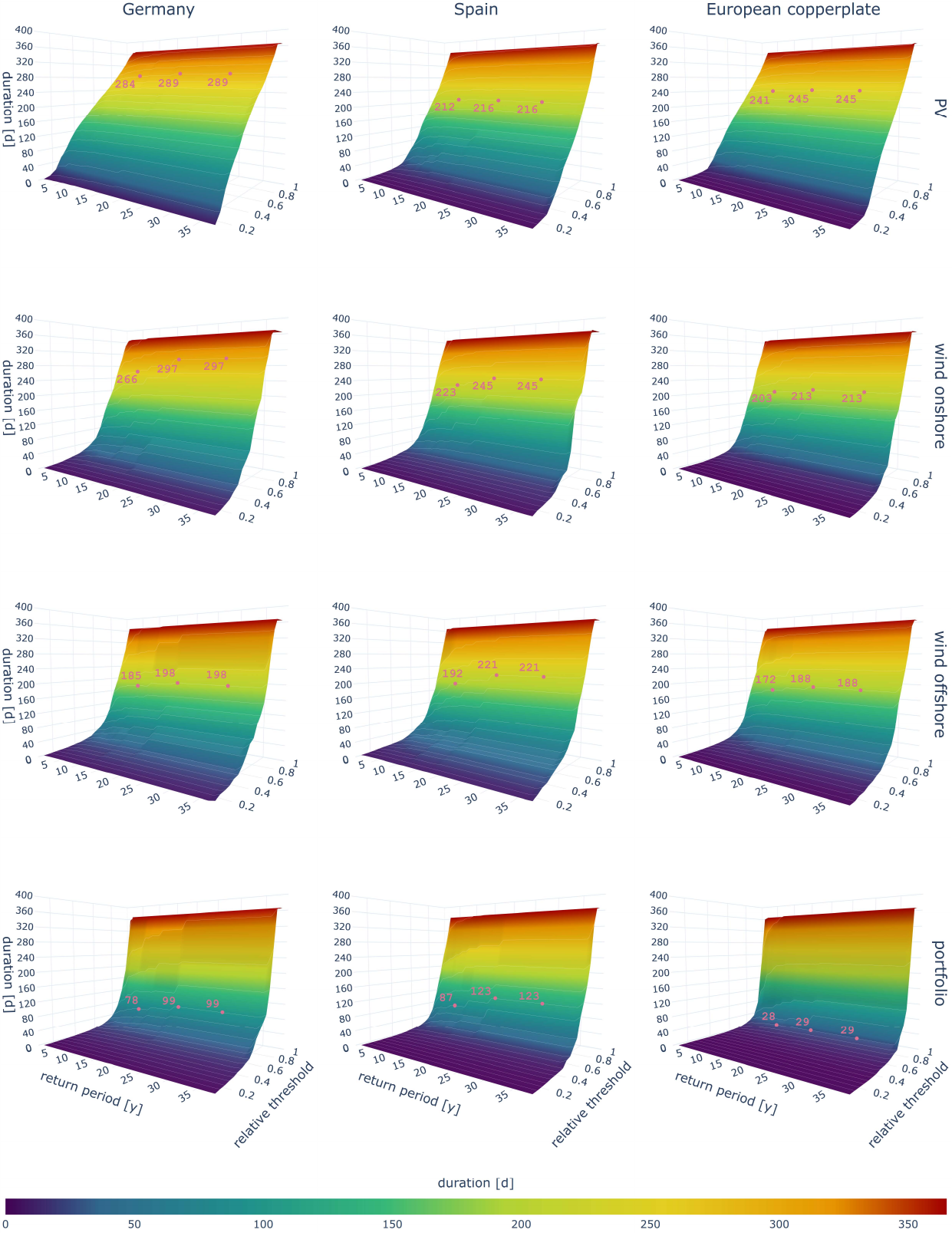}
\caption{Example of return period-duration distributions of rare drought events. \\ The Figure shows events with an average annual frequency below 1 across all investigated thresholds $\tau_i$ with $i \in [0.1, ..., 1]$. The contour lines along the return period axis represent the threshold-specific return period. The maximum duration of an event returning on average every 10, 20, and 30 years are marked for a threshold of $\tau_{0.75}$.}
\label{fig:figure_SI10}
\end{figure}

Complementary to Figure~\ref{fig:figure_3}, Figure~\ref{fig:figure_SI10} shows the return period-duration distributions of rare drought events for all employed thresholds.

\subsection{Supplementary Note 5: Additional illustrations of maximum drought durations}\label{ssec:sup_inf_max_duration}

\begin{figure}[p!]
\centering
\noindent\includegraphics[width=.97\linewidth,height=\textheight, keepaspectratio]{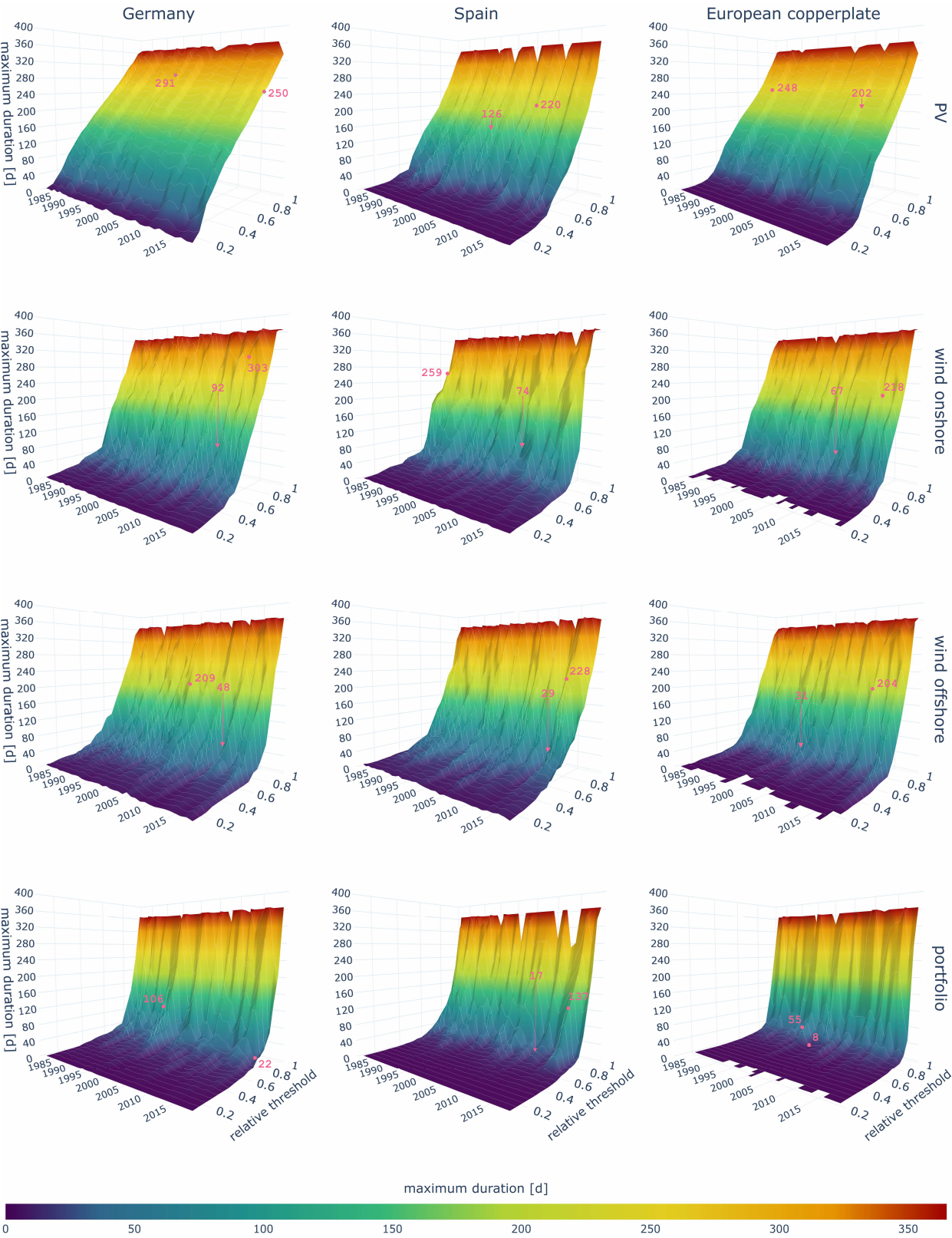}
\caption{Examples of maximum duration of single drought events per year. \\ The Figure shows the most extreme duration of single drought events for each year and across all investigated thresholds $\tau_i$ with $i \in [0.1, ..., 1]$. The contour lines represent threshold-specific maximum duration. White space indicates the absence of droughts for given thresholds in the data. The events with the highest and lowest duration across all years are marked for a threshold $\tau_{0.75}$. Arrows indicate values that are hidden in valleys of the distribution plane.}
\label{fig:figure_SI11}
\end{figure}

\begin{figure}[H]
\centering
\subfloat[Longest-lasting solar \ac{PV} droughts. Maximum droughts duration in South European countries is lower than in Northern Europe. In the European copperplate scenario (CP), extreme durations in Northern Europe can be mitigated through geographical balancing with Southern Europe (\textit{balancing effect}).\label{fig:figure_SI12a}]
    {{\includegraphics[width=\textwidth]{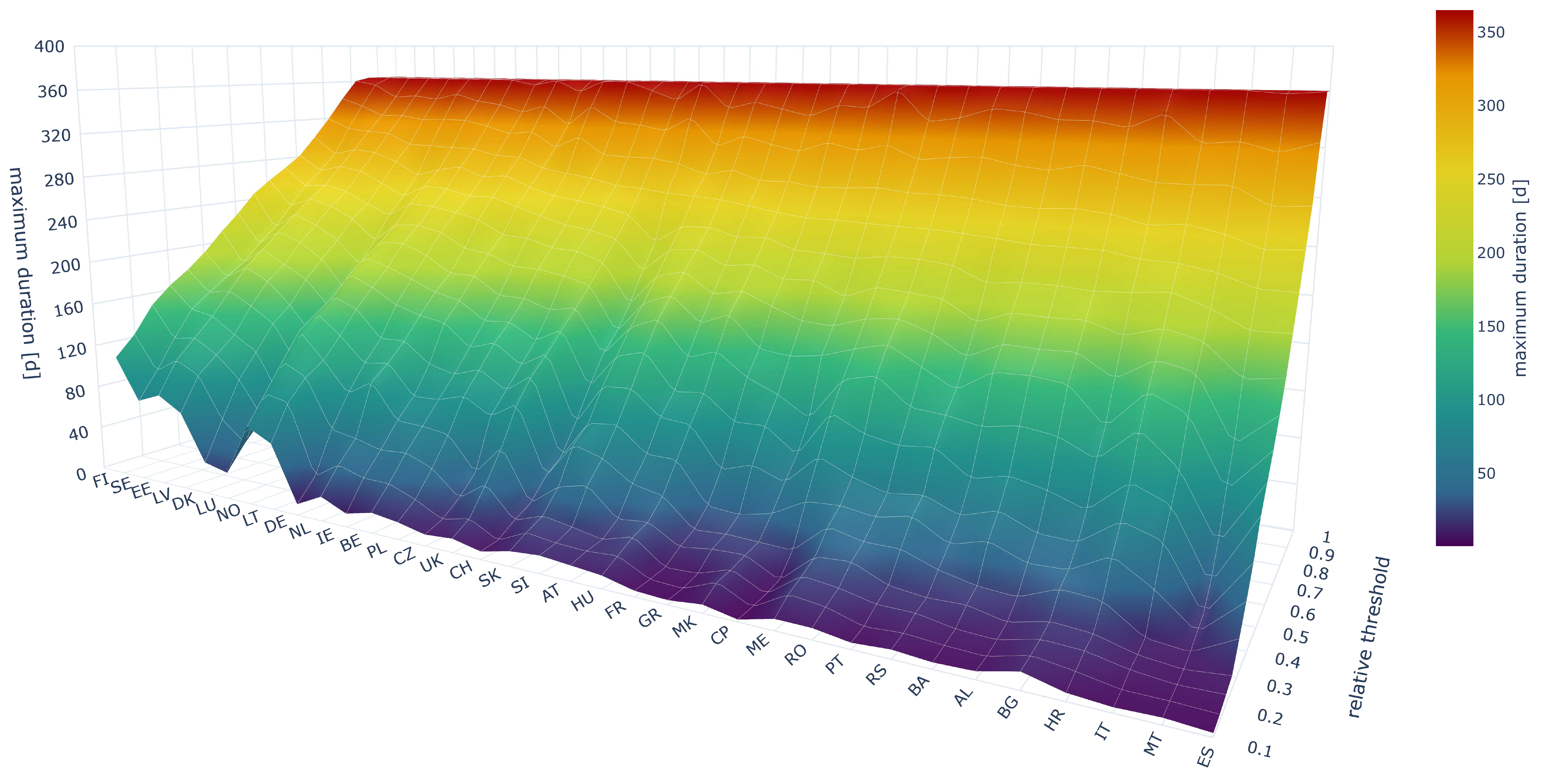}}}
\quad
\subfloat[Longest-lasting \ac{VRE} portfolio droughts. For thresholds $\tau_{i < 1.0}$, portfolio droughts are generally shorter than single-technology ones (\textit{portfolio effect}).\label{fig:figure_SI12b}]
    {{\includegraphics[width=\textwidth]{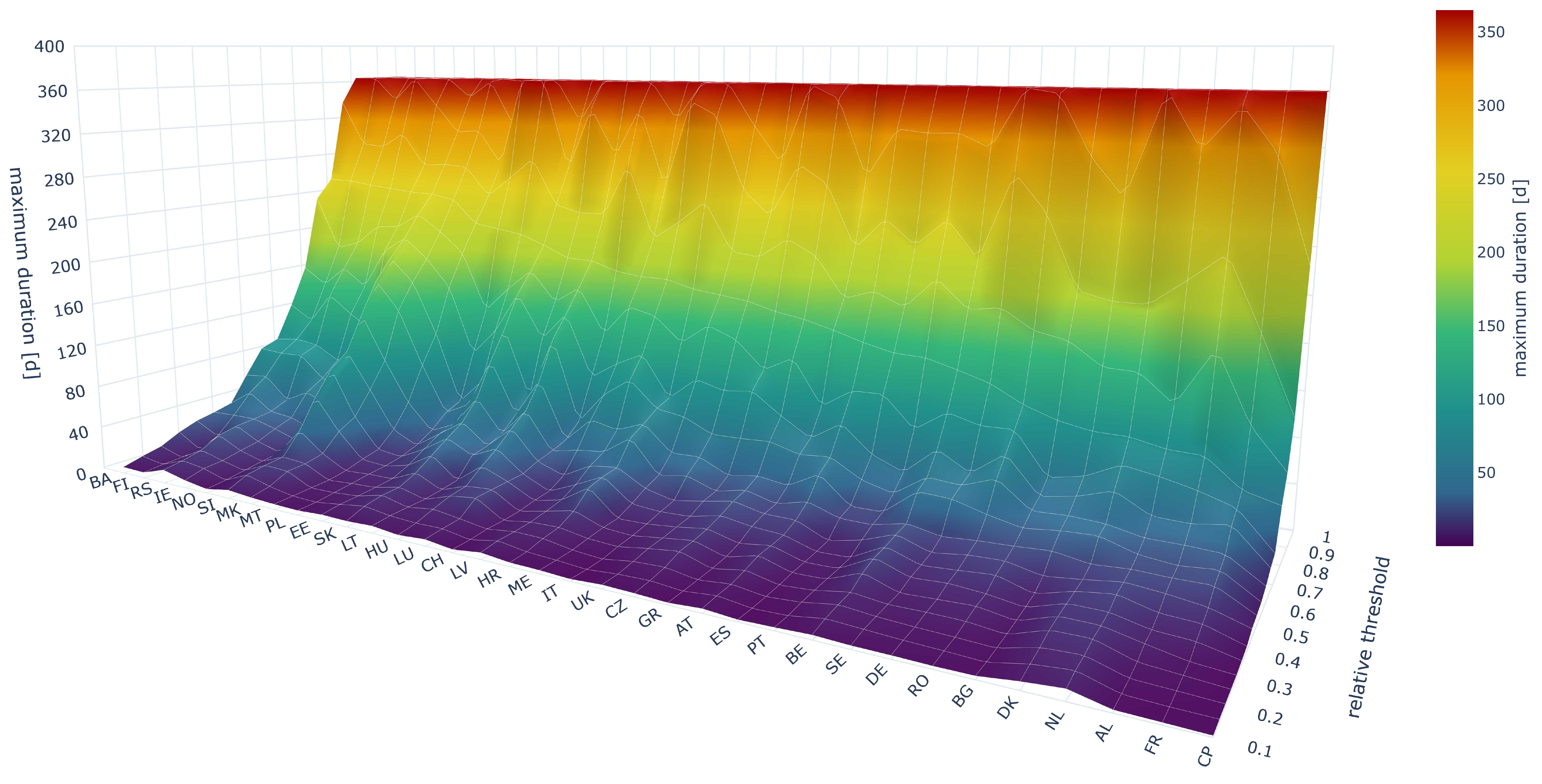}}}\\
\caption{Maximum duration of single drought event of all years for each country. \\ The Figure shows identified extreme events across all investigated thresholds $\tau_i$ with $i \in [0.1, ..., 1]$. The latter are sorted in descending order from left to right according to the longest duration for a threshold $\tau_{0.75}$.}%
\label{fig:figure_SI12}%
\end{figure}

\begin{figure}[H]
\centering
\subfloat[Longest-lasting onshore wind droughts.\label{fig:figure_SI13a}]
    {{\includegraphics[width=\textwidth]{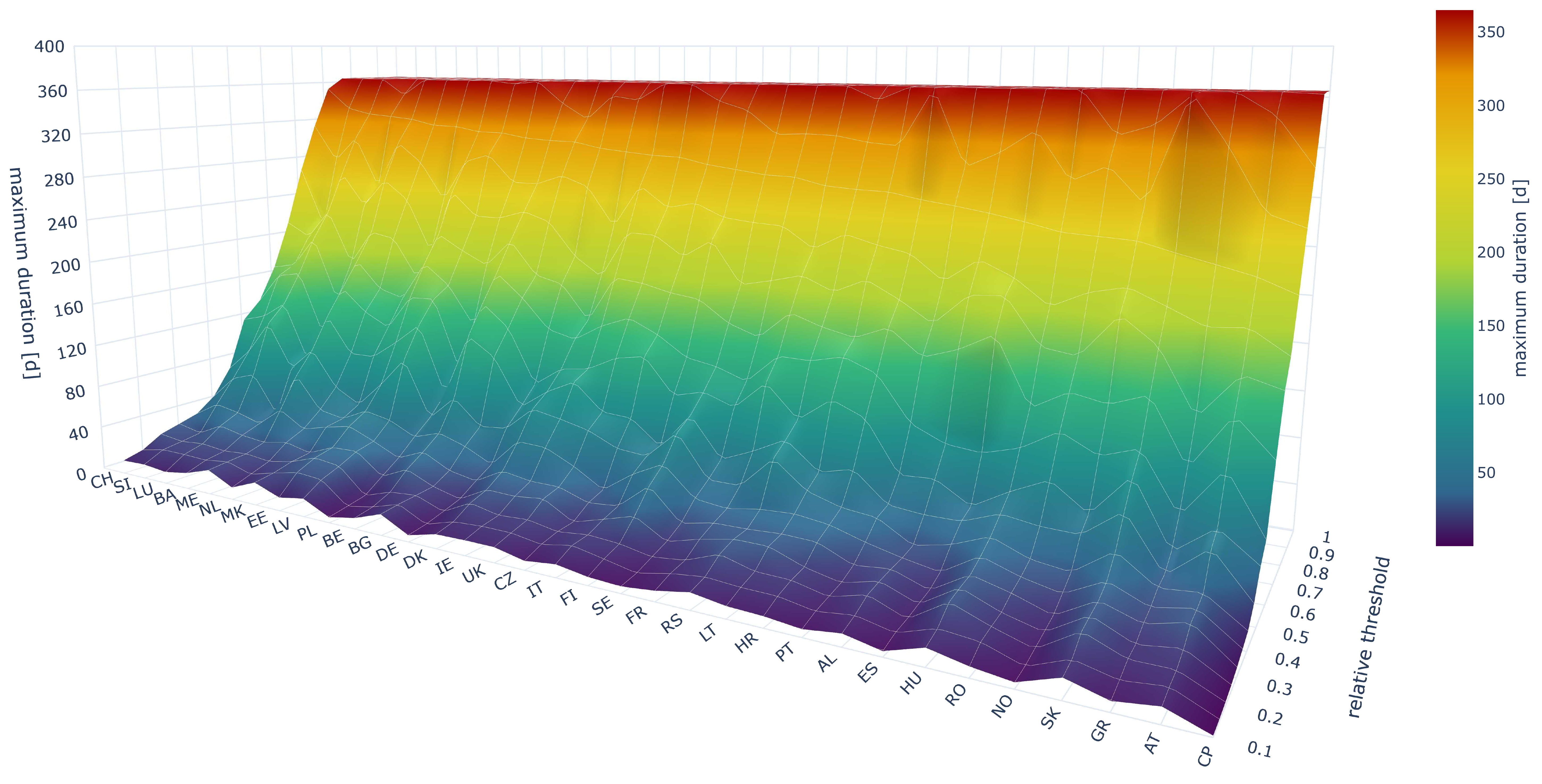}}}
\quad
\subfloat[Longest-lasting offshore wind droughts.\label{fig:figure_SI13b}]
    {{\includegraphics[width=\textwidth]{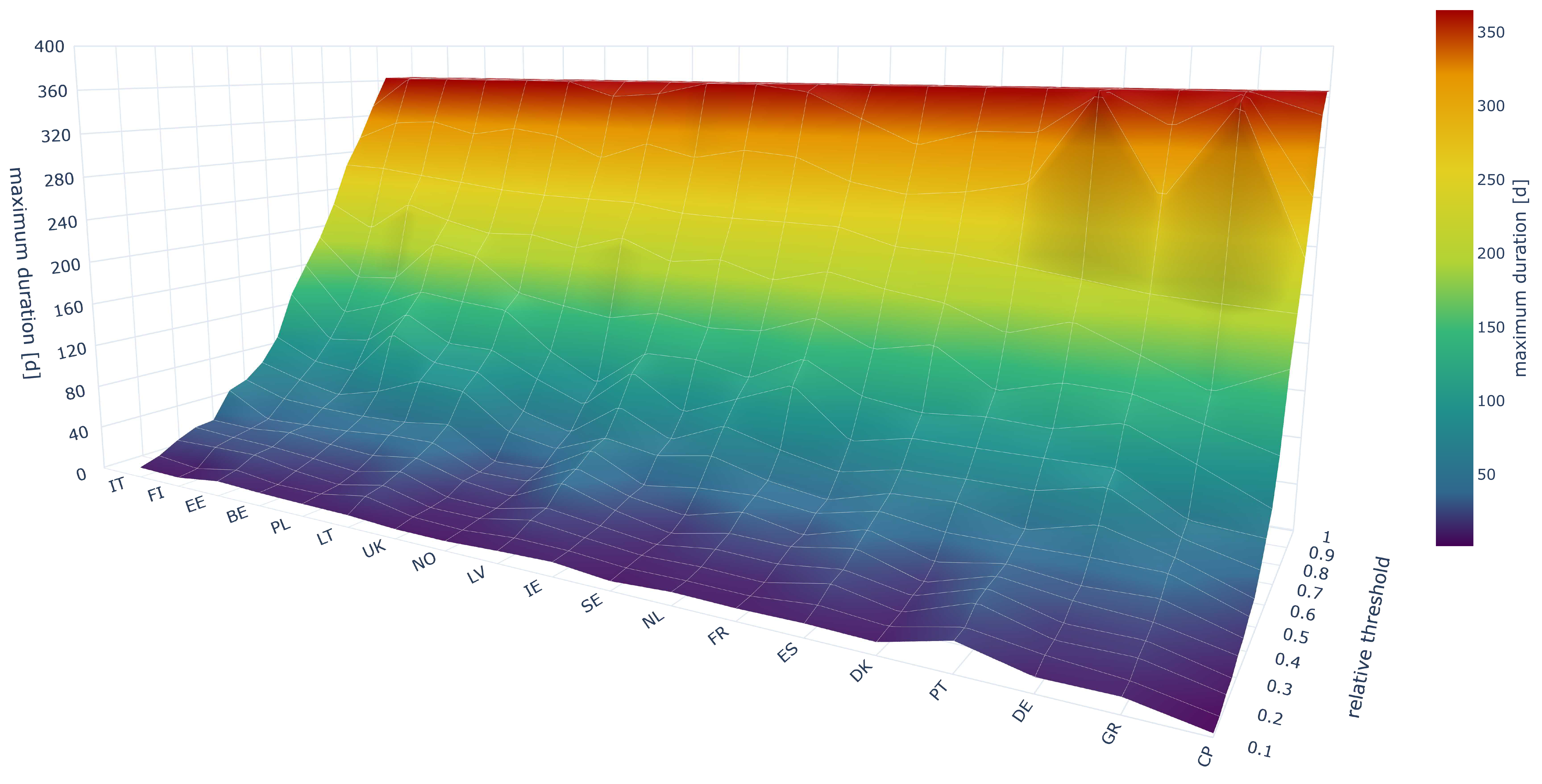}}}\\
\caption{Maximum duration of single drought event of all years for each country. \\ The Figure shows identified extreme events across all investigated thresholds $\tau_i$ with $i \in [0.1, ..., 1]$. The latter are sorted in descending order from left to right according to the longest duration for a threshold $\tau_{0.75}$. In the European copperplate scenario (CP), unconstrained geographical balancing mitigates the most extreme drought duration (\textit{balancing effect}).}%
\label{fig:figure_SI13}%
\end{figure}

\begin{figure}[H]
\centering
\noindent\includegraphics[width=\linewidth,height=\textheight, keepaspectratio]{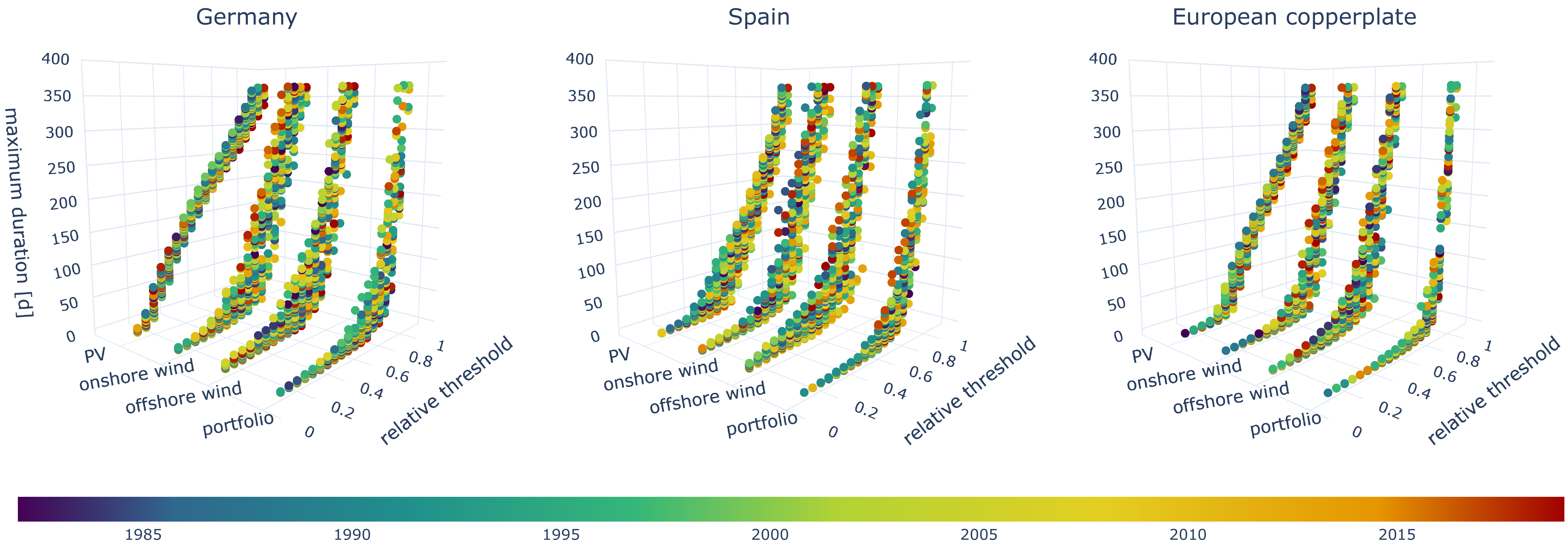}
\caption{Maximum duration of single drought events per year. \\ Most extreme duration of single drought events per year and across all investigated thresholds $\tau_i$ with $i \in [0.1,..., 1]$. Higher thresholds find long-lasting events. The year with the most extreme event duration varies across thresholds. The difference between years increases for increasing thresholds before it decreases again for very high thresholds. The ranking of years changes across thresholds. In general, combining technologies (\textit{portfolio effect}) and countries (\textit{balancing effect}) reduces the most extreme event duration.}
\label{fig:figure_SI14}
\end{figure}

Figure~\ref{fig:figure_SI11} shows the longest droughts obtained from the data for each year and threshold for Germany, Spain, and the European copperplate scenario. Figures~\ref{fig:figure_SI12} and \ref{fig:figure_SI13} show the maximum duration of single drought events across all years for each country and threshold. Figure~\ref{fig:figure_SI14} illustrates the inter-annual variability of the maximum drought duration for each threshold and technology (portfolio).

\subsection{Supplementary Note 6: Additional information on portfolio and balancing effects}\label{ssec:sup_inf_portfolio_balancing}

\begin{table}[htbp]
  \centering
  \caption{Portfolio and balancing effect. \\ Change of the maximum drought duration as unweighted averages over all investigated thresholds $\tau_i$ with $i \in [0.1,..., 1.0]$ with~$0.05$ increments in percent. The average portfolio and balancing effects are weighted averages considering the theoretical generation potential of each country.}
  \scriptsize
  \begin{tabular*}{\textwidth}{@{\extracolsep{\fill}}lccccrcccc@{}}
    \toprule
     & & \multicolumn{3}{c}{\textbf{Portfolio effect}} & & \multicolumn{4}{c}{\textbf{Balancing effect}} \\
     & & \multicolumn{3}{c}{\ac{VRE} portfolio compared to single technologies} & & \multicolumn{4}{c}{European copperplate (CP) compared to single countries} \\
    \cmidrule{3-5} \cmidrule{7-10}
    region & \phantom{a} & solar \ac{PV} & onshore wind & offshore wind & \phantom{a} & solar \ac{PV} & onshore wind & offshore wind & portfolio \\
    \cmidrule{1-1} \cmidrule{3-5} \cmidrule{7-10}
    AL    &       & -59   & -58   &       &       & 19    & -43   &       & -64 \\
    AT    &       & -61   & -46   &       &       & -10   & -38   &       & -70 \\
    BA    &       & -11   & -30   &       &       & 7     & -52   &       & -80 \\
    BE    &       & -71   & -59   & -48   &       & -37   & -57   & -43   & -70 \\
    BG    &       & -64   & -69   &       &       & 34    & -51   &       & -58 \\
    CH    &       & -58   & -55   &       &       & -10   & -51   &       & -72 \\
    CZ    &       & -67   & -51   &       &       & -32   & -50   &       & -71 \\
    DE    &       & -77   & -60   & -54   &       & -44   & -53   & -37   & -65 \\
    DK    &       & -77   & -60   & -44   &       & -45   & -50   & -26   & -65 \\
    EE    &       & -68   & -49   & -31   &       & -48   & -58   & -41   & -77 \\
    ES    &       & -61   & -58   & -63   &       & 57    & -34   & -28   & -53 \\
    FI    &       & -66   & -32   & -20   &       & -50   & -49   & -34   & -79 \\
    FR    &       & -72   & -68   & -66   &       & -2    & -48   & -33   & -50 \\
    GR    &       & -59   & -38   & -54   &       & 30    & 2     & -27   & -61 \\
    HR    &       & -51   & -53   &       &       & -5    & -50   &       & -72 \\
    HU    &       & -49   & -28   &       &       & -19   & -39   &       & -77 \\
    IE    &       & -53   & -28   & -5    &       & -37   & -48   & -28   & -79 \\
    IT    &       & -50   & -59   & -56   &       & 33    & -52   & -46   & -68 \\
    LT    &       & -71   & -43   & -39   &       & -46   & -48   & -37   & -74 \\
    LU    &       & -69   & -57   &       &       & -39   & -59   &       & -72 \\
    LV    &       & -70   & -53   & -36   &       & -47   & -58   & -37   & -74 \\
    ME    &       & -58   & -64   &       &       & -7    & -60   &       & -72 \\
    MK    &       & -11   & -11   &       &       & 20    & -49   &       & -81 \\
    MT    &       & 0     &       &       &       & 121   &       &       & -78 \\
    NL    &       & -74   & -59   & -48   &       & -42   & -56   & -41   & -68 \\
    NO    &       & -65   & -8    & 2     &       & -46   & -35   & -18   & -78 \\
    PL    &       & -66   & -46   & -37   &       & -38   & -52   & -32   & -73 \\
    PT    &       & -60   & -56   & -59   &       & 32    & -31   & -31   & -62 \\
    RO    &       & -64   & -58   &       &       & 1     & -43   &       & -62 \\
    RS    &       & -12   & -23   &       &       & 6     & -47   &       & -79 \\
    SE    &       & -76   & -53   & -46   &       & -47   & -41   & -24   & -65 \\
    SI    &       & -22   & 2     &       &       & -21   & -55   &       & -83 \\
    SK    &       & -49   & -22   &       &       & -27   & -47   &       & -79 \\
    UK    &       & -70   & -59   & -47   &       & -37   & -54   & -35   & -69 \\    
    \cmidrule{1-1} \cmidrule{3-5} \cmidrule{7-10}
    CP    &       & -80   & -45   & -70   &       &       &       &       &  \\
    \cmidrule{1-1} \cmidrule{3-5} 
    \textbf{Average} &       & \textbf{-64}   & \textbf{-52}   & \textbf{-47}   &       & \textbf{-1}   & \textbf{-46}   & \textbf{-34}   & \textbf{-65} \\
    \bottomrule
    \end{tabular*}%
  \label{tab:table_1}%
\end{table}

\newpage

\subsection{Supplementary Note 7: Additional illustrations of most extreme drought events}\label{ssec:sup_inf_extreme_events}

\begin{figure}[H]
\centering
\noindent\includegraphics[width=\linewidth,keepaspectratio]{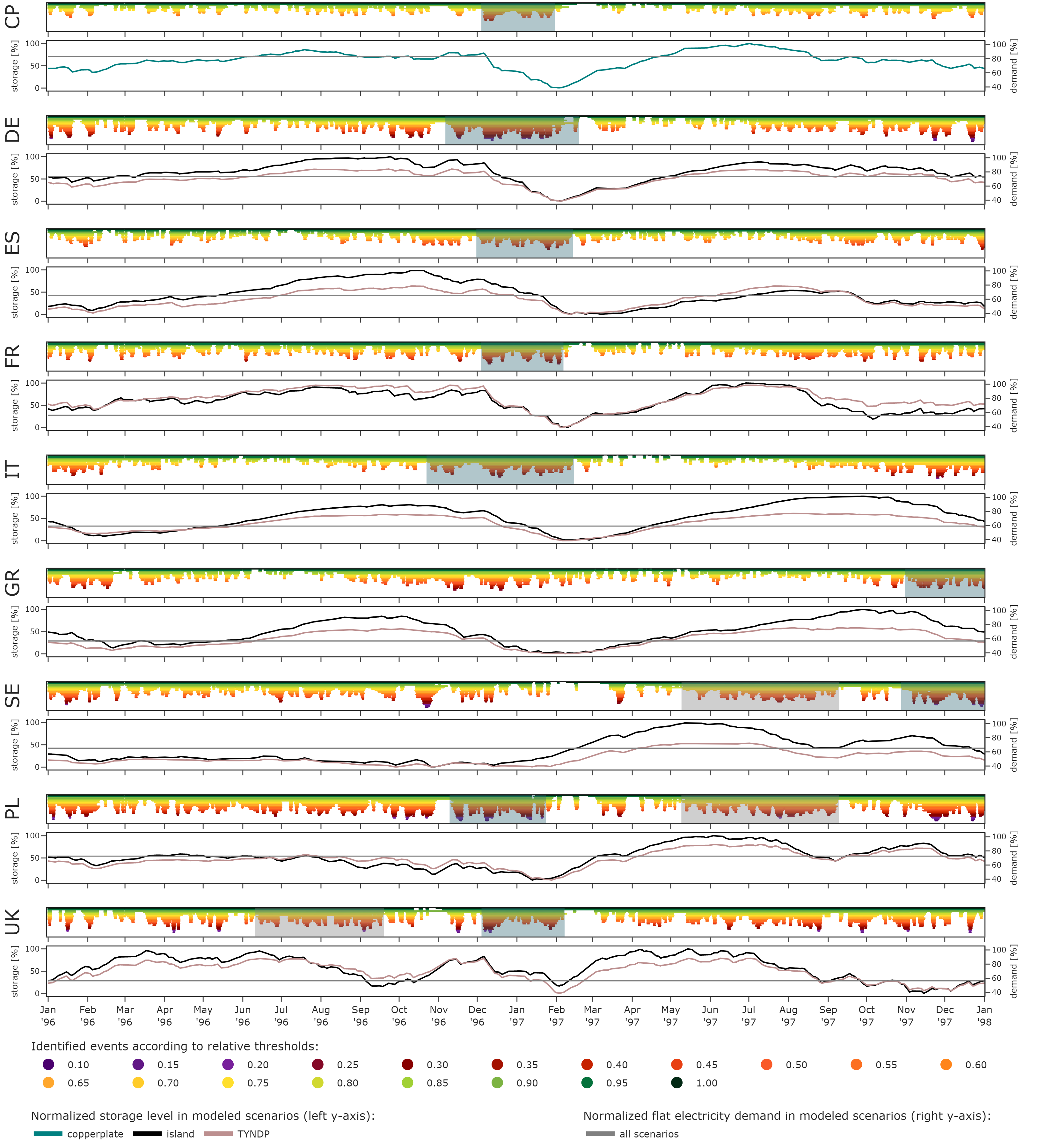}
\caption{Most extreme drought events in 1996/97 of selected countries. \\ Identified most extreme drought events in 1996/97 occurring in winter (teal boxes). For countries in which the most extreme drought events occur in summer, these are additionally shown (gray boxes). The dates shown on the bottom axis correspond to all panels. For each region, portfolio drought occurrences lasting longer than one day for stacked, color-coded thresholds (upper panel) as well as exogenous flat demand profiles and optimized storage levels across three modeled interconnection scenarios (lower panel) are displayed. Note that the major storage discharging periods now perfectly coincide with the most extreme events identified by the drought mass metric in all countries, including the summer-time droughts in Sweden, Poland, and the United Kingdom.}
\label{fig:figure_SI15}
\end{figure}

Figure~\ref{fig:figure_SI15} illustrates the most extreme drought events and optimal storage use in 1996/97 for selected regions assuming flat demand profiles, i.e.,~eliminating any diurnal or seasonal demand variability.

\end{document}